\documentclass[a4paper,11pt]{article}
\pdfoutput=1

\usepackage{jheppub}
\usepackage{natbib}
\usepackage[dvipsnames]{xcolor}

\usepackage[utf8]{inputenc}
\usepackage{diagbox}

\usepackage{todonotes}
\usepackage{enumitem}
\usepackage{amssymb,amsmath,amsbsy,mathtools,amsthm}
\usepackage{braket}
\newcommand{\tr}{\mathrm{tr}}
\usepackage{hyperref}

\usepackage{float, caption}
\usepackage{booktabs}

\usepackage{stmaryrd}

\newtheorem{lemma}{Lemma}

\newtheorem{conj}{Conjecture}

\newcommand{\ul}[1]{{\underline{#1}}}

\usepackage{graphicx}

\title{Entanglement cohomology for GHZ and W states}

\author[a,b]{Christian Ferko}
\author[a]{Keiichiro Furuya}

\affiliation[a]{Department of Physics, Northeastern University, Boston, MA, 02115 USA}
\affiliation[b]{The NSF Institute for Artificial Intelligence and Fundamental Interactions}

\emailAdd{c.ferko@northeastern.edu}
\emailAdd{k.furuya@northeastern.edu}

\abstract{Entanglement cohomology assigns a graded cohomology ring to a multipartite pure state, providing homological invariants that are stable under local unitaries and characterize inequivalent patterns of entanglement. In this work we derive exact expressions for the dimensions of these cohomology groups in two canonical entanglement classes, generalized GHZ and W states on an arbitrary number of parties and local Hilbert space dimensions, thus proving conjectures of \cite{Mainiero:2019enr}. Using the additional structure of the Hodge star and wedge product operations, we propose two new classes of local unitary invariants: the spectrum of the natural Laplacian acting on entanglement $k$-forms, and the intersection numbers obtained from wedge products of representatives for cohomology classes. We present numerical experiments which investigate these invariants in particular states, suggesting that they may provide useful quantities for describing multipartite entanglement.}

\makeatletter
\gdef\@fpheader{}
\makeatother
\begin{document}

\maketitle

\section{Introduction}\label{sec:intro}

Entanglement is among the most fundamental features which distinguish quantum mechanics from classical physics. It is also an intrinsically \emph{obstruction-theoretic} phenomenon, in the following sense. We say that a pure state in a multipartite Hilbert space,
\begin{align}
    \ket{\psi} \in \mathcal{H}_{\ul{1}} \otimes \ldots \otimes \mathcal{H}_{\ul{n}} \, ,
\end{align}
is a \emph{product state} if it can be written as a tensor product of states in each space $\mathcal{H}_{\ul{i}}$,
\begin{align}
    \ket{\psi} = \ket{\psi_{\ul{1}}} \otimes \ldots \otimes \ket{\psi_{\ul{n}}} \, , \qquad \ket{\psi_{\ul{i}}} \in \mathcal{H}_{\ul{i}} \, ,
\end{align}
and we say that $\ket{\psi}$ is \emph{entangled} otherwise. More generally, for mixed states, one says that a density matrix $\rho$ represents a \emph{separable} state if it can be written as a convex combination of product states, and $\rho$ defines an entangled state otherwise; separable states exhibit classical correlations but no genuine quantum entanglement.\footnote{However we note that, for mixed states, separability does not rule out other forms of nonclassical correlations such as quantum discord. See, for instance, \cite{OllivierZurek2001Discord,HendersonVedral2001Correlations,ModiEtAl2012RMPDiscord,DakicVedralBrukner2010Discord}.} Although we will focus on pure states in this work, we note that both of these definitions of entanglement are phrased in terms of what entangled states are \emph{not}. Entanglement is precisely the obstruction from realizing a pure state as a product, or a mixed state as a separable state.

A natural mathematical framework for studying obstruction-theoretic questions is that of homological algebra. Perhaps the most familiar example is that of de Rham cohomology on a smooth manifold $M$. Consider the spaces $\Omega^k ( M )$ of differential forms on $M$, along with the exterior derivative operator $d$ which maps $k$-forms to $(k+1)$-forms. Every closed differential form is locally exact, by the Poincar\'e lemma. The de Rham cohomology groups 
\begin{align}
    H_{\mathrm{dR}}^k ( M ) = \frac{\mathrm{ker} \left( d^k : \Omega^k ( M ) \to \Omega^{k+1} ( M ) \right) }{\mathrm{im} \left( d^{k-1} : \Omega^{k-1} ( M ) \to \Omega^k ( M ) \right) }
\end{align}
measure the obstruction from promoting a locally exact form to a globally exact form.

An even simpler example\footnote{This example is called ``cohomology on a bumper sticker'' in \cite{lim2020hodge}, to which we refer the reader for a pedagogical discussion of cohomology, Hodge theory, and Laplacians on graphs.} can be stated using only linear algebra. Consider three vector spaces $V_0, V_1, V_2$ together with linear maps $N : V_0 \to V_1$, $M : V_1 \to V_2$ obeying
\begin{align}\label{MN_zero}
    M N = 0 \, .
\end{align}
If a vector $v$ can be written as $v = N w$ for some $w$, then certainly $M v = M N w = 0$, so
\begin{align}
    \mathrm{im} ( N ) \subseteq \mathrm{ker} ( M ) \, .
\end{align}
However, given only the knowledge that $M v = 0$, one is obstructed from concluding that $v = N w$ for some $w$, since there may be vectors in the kernel of $M$ but not in the image of $N$. This obstruction is again characterized by the cohomology
\begin{align}\label{matrix_cohomology}
    H = \frac{\ker \left( M : V_1 \to V_2 \right)}{\mathrm{im} \left( N : V_0 \to V_1 \right) } \, .
\end{align}
In some cases, one can think of $V_0$ as ``local data'' (e.g. values attached to patches or vertices), $V_1$ as ``overlap data'' (values attached to pairwise overlaps or edges), and $V_2$ as higher-overlap constraints (triple overlaps, cycles, faces, and so on). A vector $v \in V_1$ is viewed as a choice of data on overlaps, and the condition $M v = 0$ expresses that this overlap data is locally consistent: it passes all constraints encoded by $M$. If, in addition, there exists a $w \in V_0$ with $v = N w$ then the overlap data $v$ actually arises from gluing together some underlying choice of local data $w$. However, if the cohomology (\ref{matrix_cohomology}) is non-trivial, there are overlap configurations that are locally consistent ($v \in \ker(M)$) but which cannot be obtained by gluing any global choice of local data ($v \notin \mathrm{im}(N)$). Thus even in this finite-dimensional linear algebra setting, cohomology can be interpreted as measuring the failure of locally consistent data to come from a single globally defined object.

Such cohomological questions -- attempting to characterize when one is prevented from gluing local data together to form a consistent global object -- have a very similar flavor to the problem of understanding quantum entanglement. Owing to this suggestive similarity, in recent years several works have applied cohomological techniques in order to describe entanglement in quantum mechanical systems. For instance, in \cite{Hamilton:2023klg}, the authors proposed a framework for measuring entanglement using the machinery of persistent homology, which was later extended in \cite{Banka:2025yds}. Another approach was explained in the recent work \cite{Ikeda:2025mgj}, which employed a sheaf-theoretic strategy, organizing states and entanglement witnesses into a presheaf and characterizing obstructions to gluing local data using \v{C}ech classes.

In this work, we will build upon a third proposal that was originally introduced in \cite{Mainiero:2019enr}, and which we refer to as \emph{entanglement cohomology}.\footnote{This approach shares some conceptual similarities with the information cohomology that was studied in \cite{e17053253,Vigneaux2017InformationSA}. See also \cite{10.1007/978-3-031-38271-0_25} for a related homological approach to classical (as opposed to quantum) probability.} The idea of this formalism, which will be reviewed in greater detail in Section \ref{sec:review}, is to measure entanglement using a certain obstruction theory involving ``gluing'' of operators which act on Hilbert spaces associated with reduced density matrices arising from a global multipartite pure state.\footnote{We describe a global pure state interchangeably as a ket $\ket{\psi}$ or a pure state density matrix $\rho=\ket{\psi}\bra{\psi}$.} One defines spaces of ``entanglement $k$-forms'', which we write as $\Omega^k ( \rho )$ by analogy with the notation for the space of differential forms on a manifold, along with ``exterior derivative'' operators:
\begin{align}
    d^k : \Omega^k ( \rho ) \to \Omega^{k+1} ( \rho ) \, .
\end{align}
As we will see, the spaces $\Omega^k ( \rho )$ are simply tuples of finite-dimensional matrices. Therefore, every such space $\Omega^k ( \rho )$ can be viewed as a finite-dimensional vector space, and the derivative operators $d^k$ are matrices or linear maps which are defined to obey
\begin{align}
    d^{k+1} d^k = 0 \, ,
\end{align}
for all $k$. This is precisely the setting of the linear-algebraic cohomology that we mentioned around equation (\ref{MN_zero}), with $M = d^{k+1}$ and $N = d^k$. The corresponding entanglement cohomology groups will therefore be defined as in (\ref{matrix_cohomology}) for each $k$.

Despite the fact that elements of $\Omega^k ( \rho )$ are mere matrices -- and not, for instance, differential forms on a manifold -- it may come as a surprise that the structure of entanglement forms shares many properties with that of differential forms defined on a manifold with a Riemannian metric \cite{Ferko:2024swt}. For instance, just as one can define a Hodge star operation
\begin{align}
    \ast : \Omega^k ( M ) \to \Omega^{n - k} ( M ) 
\end{align}
on an $n$-manifold with a metric, there is an analogous Hodge star that maps entanglement $k$-forms to entanglement $(n-k)$-forms, in the setting of pure states on $n$-partite Hilbert spaces. Likewise, one can define a wedge product of entanglement forms, and therefore a natural inner product which takes the schematic form $\langle \omega , \eta \rangle = \tr ( \omega \wedge \ast \eta )$. Continuing to take inspiration from Hodge theory, we can define the operator $\delta$ which is adjoint to $d$ with respect to this inner product, and from there construct the Laplacian
\begin{align}\label{lap_defn}
    \Delta = d \delta + \delta d \, ,
\end{align}
which we will revisit in Section \ref{sec:lap}.

This rich structure suggests that one might learn more about entanglement in finite-dimensional quantum systems by further exploiting this analogy, computing quantities that are motivated by geometrical considerations and studying their physical interpretations. Such quantities carry more information than the \emph{Betti numbers},
\begin{align}\label{betti_defn}
    b_k ( \rho_{\ul{1} \ldots \ul{n}} ) = \mathrm{dim} \left( H^k ( \rho_{\ul{1} \ldots \ul{n}} ) \right) \, ,
\end{align}
associated with a pure state. The purpose of this work is to continue this exploration of the use of entanglement cohomology as a tool for characterizing multipartite entanglement, both by establishing new results about the dimensions (\ref{betti_defn}) and by introducing new geometrically-inspired entanglement measures. More specifically, our goals are twofold:

\begin{enumerate}[label = (\roman*)]

    \item\label{intro_prove} We will rigorously prove conjectural closed-form expressions, put forward in \cite{Mainiero:2019enr} based on substantial numerical evidence, for the dimensions (\ref{betti_defn}) of the entanglement cohomology groups in two canonical classes of states, namely generalized GHZ and W states. While filling a gap in the literature, the proof technique that we develop in order to establish these results also reveals an interesting connection between entanglement cohomology and ordinary simplicial homology. This method might be used to prove results about Betti numbers for other classes of quantum states, and it serves to further highlight which characteristics of entanglement (e.g. GHZ versus W type entanglement patterns) are captured by the mere \emph{dimensions} of cohomologies, to be contrasted with properties that can only be probed by computing other quantities.

    \item\label{intro_invariants} We will introduce two new classes of local unitary (LU) invariants, the spectrum of the entanglement Laplacian (\ref{lap_defn}) and the collection of \emph{intersection numbers}, that can be computed from any multipartite pure state by leveraging the geometrical character of entanglement $k$-forms. The first of these, the spectrum of $\Delta$, contains strictly more information than the Betti numbers $b_k$: the dimension of $H^k$ is identical to the number of zero eigenvalues of $\Delta : \Omega^k \to \Omega^k$, while the non-zero eigenvalues furnish us with new invariants. Likewise, we will show that double intersection numbers contain information about Schmidt coefficients of the parent pure state across bipartitions -- additional data which is not contained in the dimensions of the cohomologies themselves -- while higher intersection numbers capture still more information about the multipartite entanglement structure. We will also present the results of numerical experiments which investigate these new LU invariants in certain classes of states.
\end{enumerate}

The study of entanglement cohomology is still quite young, and many basic questions remain to be explored. It is our hope that both sets of results \ref{intro_prove} and \ref{intro_invariants} will spur further progress in this area, leading to the development of additional mathematical tools that might eventually allow us to chart the space of possible patterns of entanglement in finite-dimensional quantum systems and thus better understand quantum information.

The structure of this paper is as follows. In Section \ref{sec:review}, we review the aspects of entanglement cohomology which are relevant for later parts of this work. Section \ref{sec:proofs} presents the proofs of two conjectures of \cite{Mainiero:2019enr} proposing exact, closed-form expressions for the dimensions of entanglement cohomologies of generalized GHZ and W states. In Section \ref{sec:invariants}, we define novel, geometrically-inspired local unitary invariants that can be constructed from the entanglement $k$-forms associated with a generic pure state, and study the structure of these new quantities in several examples. Section \ref{sec:conclusion} summarizes our conclusions and describes a few areas for future research. We have relegated some standard material concerning simplicial homology to Appendix \ref{app:simp} for the benefit of the unfamiliar reader.

\section{Generalities on Entanglement Cohomology}\label{sec:review}

In order to make the present work self-contained, and to fix our conventions, we will begin by reviewing the formalism of entanglement cohomology. Although this subject was first introduced in \cite{Mainiero:2019enr}, we will use notation which more closely follows that of \cite{Ferko:2024swt}.

We will always work in a finite tensor product of finite-dimensional Hilbert spaces,
\begin{align}\label{general_hilbert_space}
    \mathcal{H}_{\ul{1} \ldots \ul{n}} = \bigotimes_{\ul{i} = \ul{1}}^{\ul{n}} \mathcal{H}_{\ul{i}} = \mathcal{H}_{\ul{1}} \otimes \ldots \otimes \mathcal{H}_{\ul{n}} \, .
\end{align}
Indices which label one of the $n$ subsystems of the total Hilbert space will be decorated with an underline, to distinguish them from other integer labels that may appear. For instance, if $\mathcal{H}_{\ul{1}}$ is $2$-dimensional, we might write basis states as $\ket{0_{\ul{1}}}$ and $\ket{1_{\ul{1}}}$, where the non-underlined integers refer to the states in the computational basis. When no confusion is possible, we will omit subsystem labels and simply write $\ket{0}$ and $\ket{1}$ for these basis states.

When it is necessary to speak of the dimensions of the tensor product factors $\mathcal{H}_{\ul{i}}$, we will denote the local dimensions by $d_{\ul{i}} = \dim \left( \mathcal{H}_{\ul{i}} \right)$, so that the total dimension is
\begin{align}
    \dim \left( \mathcal{H}_{\ul{1} \ldots \ul{n}} \right) = \prod_{\ul{i} = \ul{1}}^{\ul{n}} d_{\ul{i}} \, .
\end{align}
The starting point for defining the entanglement cochain complex is to select a pure state
\begin{align}\label{pure_state_defn}
    \ket{\psi_{\ul{1} \ldots \ul{n}}} \in \mathcal{H}_{\ul{1} \ldots \ul{n}} \, ,
\end{align}
to which we can associate the density matrix
\begin{align}\label{parent_rho}
    \rho_{\ul{1} \ldots \ul{n}} = \ket{\psi_{\ul{1} \ldots \ul{n}}} \bra{\psi_{\ul{1} \ldots \ul{n}}} \, .
\end{align}
Although it is also possible to directly study entanglement cohomology for mixed states, the resulting structure does not enjoy the same mathematical properties as for pure states. For instance, there is no well-defined Hodge star operation acting on mixed state entanglement $k$-forms, which precludes us from building the entanglement Laplacian to be introduced in Section \ref{sec:lap}. For this reason, we will take the view that mixed states should always be studied by first purifying and then computing invariants using the entanglement cohomology of the parent pure states. We therefore focus entirely on pure states in what follows.

Given a parent density matrix $\rho_{\ul{1} \ldots \ul{n}}$ as in (\ref{parent_rho}), its various reduced density matrices -- obtained by tracing out some number of tensor product factors -- will be denoted with the same symbol $\rho$ but a different number of underlined indices. For instance,
\begin{align}
    \rho_{\ul{2} \ldots \ul{n}} = \tr_{\ul{1}} \left( \rho_{\ul{1} \ldots \ul{n}} \right) \, , \qquad \rho_{\ul{3}} = \tr_{\ul{1} \ul{2} \ul{4} \ldots \ul{n}} \left( \rho_{\ul{1} \ldots \ul{n}} \right) \, ,
\end{align}
and so on. Sometimes we will use a hat to indicate that a label has been omitted, as in
\begin{align}
    \rho_{ \ul{1} \widehat{\ul{2}} \ul{3} \ldots \ul{n}} = \rho_{\ul{1} \ul{3} \ldots \ul{n}} = \tr_{\ul{2}} \left( \rho_{\ul{1} \ldots \ul{n}} \right) \, .
\end{align}
Here the partial trace is defined in the usual way. Given any linear operator
\begin{align}
    \mathcal{O} : V \otimes W \to V \otimes W
\end{align}
acting on a tensor product of vector spaces $V$ and $W$, where $\{ e_1 , \ldots, e_m \}$ is a basis for $V$ and $\{ f_1, \ldots, f_n \}$ is a basis for $W$, let $\mathcal{O}_{k \ell , i j}$ be the matrix elements of this operator with respect to the basis $\{ e_k \otimes f_\ell \}$, where $1 \leq k, i \leq m$ and $1 \leq \ell, j \leq n$. The partial trace $\tr_W \left( \mathcal{O} \right) : V \to V$ is the operator whose matrix elements with respect to the basis $\{ e_k \}$ are
\begin{align}
    \left( \tr_W \left( \mathcal{O} \right) \right)_{k, i} = \sum_{j=1}^{n} \mathcal{O}_{k j , i j} \, .
\end{align}
In addition to the lowercase underlined Latin indices that denote subsystems, we will use capital underlined Latin letters for multi-indices (collections of $k$ subsystem indices) like
\begin{align}
    \ul{I} = \left( \ul{i}_1 , \ldots, \ul{i}_k \right) \, ,
\end{align}
which we always assume have been arranged in increasing order,
\begin{align}
    \ul{i}_1 < \ldots < \ul{i}_k \, .
\end{align}
Given such a multi-index $\ul{I}$, we write $\ul{I}^C$ for its complement in the full list of subsystem indices $\ul{1} , \ldots, \ul{n}$, again taken to be sorted in increasing order.

There is a space of linear operators $\mathrm{End} \left( \mathcal{H}_{\ul{I}} \right)$ acting on each reduced Hilbert space $\mathcal{H}_{\ul{I}}$, for each length-$k$ multi-index $\ul{I}$ and for each $k = 1 , \ldots, n$, and an associated reduced density matrix $\rho_{\ul{I}}$. We write $s_{\rho_{\ul{I}}}$, or simply $s_{\ul{I}}$ for short, to denote the projection operator onto the image of this reduced density matrix (i.e. the span of its nonzero eigenvalues):
\begin{align}
    s_{\ul{I}} = \mathrm{proj}_{\mathrm{im} ( \rho_{\ul{I} } ) } \, .
\end{align}
Because the parent state is pure, $s_{\ul{1} \ldots \ul{n}}$ will generically be a rank-one projector, but the other support projections $s_{\ul{I}}$ may have different ranks.

Given an operator $\mathcal{O} : \mathcal{H}_{\ul{I}} \to \mathcal{H}_{\ul{I}}$, we denote its restriction to the image of the associated reduced density matrix with a vertical bar:
\begin{align}
    \mathcal{O} \big\vert_{\rho_{\ul{I}}} = s_{\ul{I}} \mathcal{O} s_{\ul{I}} \, .
\end{align}
Said differently, the bar notation indicates that the endomorphism $\mathcal{O} : \mathcal{H}_{\ul{I}} \to \mathcal{H}_{\ul{I}}$ has been restricted to an operator $\mathcal{O} \big\vert_{\rho_{\ul{I}}}  : \mathrm{im} ( \rho_{\ul{I}} ) \to \mathrm{im} ( \rho_{\ul{I}} )$.

The key objects of our attention are \emph{entanglement forms} which have a structure similar to that of differential forms on a manifold. There is one such space of entanglement forms $\Omega^k ( \rho_{\ul{1} \ldots \ul{n}} )$ for each integer $k = 0 , \ldots, n$. The $0$-forms are simply complex numbers,
\begin{align}
    \Omega^0 ( \rho_{\ul{1} \ldots \ul{n}} ) = \mathbb{C} \, ,
\end{align}
and the top forms are just linear operators acting on the image of $\rho_{\ul{1} \ldots \ul{n}}$,
\begin{align}
    \Omega^n ( \rho_{\ul{1} \ldots \ul{n}} ) = \mathrm{End} \left( \mathrm{im} ( \rho_{\ul{1} \ldots \ul{n}} ) \right) \cong \mathbb{C} \, ,
\end{align}
which is also one-dimensional because the parent state is pure.

For other values $0 < k < n$, an entanglement form is a tuple of operators acting on images of reduced density matrices:
\begin{align}
    \Omega^k ( \rho_{\ul{1} \ldots \ul{n}} ) = \bigtimes_{ | \ul{I} | = k } \mathrm{End} \left( \mathrm{im} ( \rho_{\ul{I}} ) \right) \, ,
\end{align}
where the Cartesian product $\bigtimes$ runs over all increasing length-$k$ multi-indices $\ul{I} = ( \ul{i}_1 , \ldots, \ul{i}_k )$.

As a simple illustration, let us consider the case of a bipartite Hilbert space $\mathcal{H}_{AB} = \mathcal{H}_A \otimes \mathcal{H}_B$. When studying Hilbert spaces with a small number of factors, such as two or three, we will sometimes use non-underlined capital early Latin letters ($A$, $B$, $C$) to label subsystems, rather than numerical labels $\ul{i}$. For a pure state $\ket{\psi_{AB}}$ with associated density matrix $\rho_{AB}$, the entanglement one-forms $\alpha_1 \in \Omega^1 ( \rho_{AB} )$ are $2$-tuples of the form
\begin{align}
    \alpha_1 = \left( \mathcal{O}_A , \mathcal{O}_B \right) \, ,
\end{align}
where
\begin{align}
    \mathcal{O}_A : \mathrm{im} ( \rho_A ) \to \mathrm{im} ( \rho_A ) \, , \qquad \mathcal{O}_B : \mathrm{im} ( \rho_B ) \to \mathrm{im} ( \rho_B ) \, .
\end{align}
For a tripartite system $\mathcal{H}_{ABC} = \mathcal{H}_A \otimes \mathcal{H}_B \otimes \mathcal{H}_C$, entanglement $1$-forms are tuples
\begin{align}
    \left( \mathcal{O}_A , \mathcal{O}_B , \mathcal{O}_C \right) \, ,
\end{align}
with three elements, where again
\begin{align}
    \mathcal{O}_A : \mathrm{im} ( \rho_A ) \to \mathrm{im} ( \rho_A ) \, , \quad \mathcal{O}_B : \mathrm{im} ( \rho_B ) \to \mathrm{im} ( \rho_B ) \, , \quad \mathcal{O}_C : \mathrm{im} ( \rho_C ) \to \mathrm{im} ( \rho_C ) \, ,
\end{align}
and $2$-forms are length-$3$ tuples of operators acting on bipartite Hilbert spaces,
\begin{align}
    \left( \mathcal{O}_{AB} , \mathcal{O}_{AC} , \mathcal{O}_{BC} \right) \, ,
\end{align}
where now
\begin{align}
    \mathcal{O}_{AB} : \mathrm{im} ( \rho_{AB} ) \to \mathrm{im} ( \rho_{AB} ) \, , \quad \mathcal{O}_{AC} : \mathrm{im} ( \rho_{AC} ) \to \mathrm{im} ( \rho_{AC} ) \, ,  \quad \mathcal{O}_{BC} : \mathrm{im} ( \rho_{BC} ) \to \mathrm{im} ( \rho_{BC} ) \, ,
\end{align}
and so on. In general, $\Omega^k ( \rho_{\ul{1} \ldots \ul{n}} )$ is a tuple of operators of length ${n \choose k}$.

In order to define the \emph{differential}, or \emph{coboundary} operator, acting on entanglement $k$-forms, it will be necessary to introduce some notation to address ordering of tensor product factors. Let $\ul{I} = ( \ul{i}_1 , \ldots, \ul{i}_k )$ be a multi-index which is a strict subset of $\ul{J} = ( \ul{j}_1 , \ldots , \ul{j}_m )$. There is a canonical identification of Hilbert spaces,
\begin{align}\label{canonical_identification}
    \mathcal{H}_{\ul{J}} \cong \mathcal{H}_{\ul{I}} \otimes \mathcal{H}_{\ul{J} \setminus \ul{I}} \, ,
\end{align}
which simply re-shuffles the tensor product factors on the right side of (\ref{canonical_identification}) to arrange them in increasing order. For instance, if $\ul{I} = ( \ul{1} , \ul{3} )$ and $\ul{J} = ( \ul{1} , \ul{2} , \ul{3} )$, then the isomorphism
\begin{align}
    \mathcal{H}_{\ul{J}} = \mathcal{H}_{\ul{1}} \otimes \mathcal{H}_{\ul{2}} \otimes \mathcal{H}_{\ul{3}} \cong  \mathcal{H}_{\ul{1}} \otimes \mathcal{H}_{\ul{3}} \otimes \mathcal{H}_{\ul{2}} = \mathcal{H}_{\ul{I}} \otimes \mathcal{H}_{\ul{J} \setminus \ul{I}} \, , 
\end{align}
permutes subsystems $\ul{2}$ and $\ul{3}$.

We denote the re-shuffling operation which implements the identification (\ref{canonical_identification}) by
\begin{align}
    \Pi_{\ul{I}, \ul{J}} : \mathcal{H}_{\ul{J}} \to \mathcal{H}_{\ul{I}} \otimes \mathcal{H}_{\ul{J} \setminus \ul{I}} \, .
\end{align}
There is a natural inclusion of $\mathrm{End} ( \mathrm{im} ( \rho_{\ul{I}} ) )$ into $\mathrm{End} ( \mathrm{im} ( \rho_{\ul{J}} ) )$, where again $\ul{I} \subset \ul{J}$, which simply ``tensors with the identity on missing sites'' belonging to $\ul{J} \setminus \ul{I}$.  Let $X \in \mathrm{End} ( \mathrm{im} ( \rho_{\ul{I}} ) )$. We define a corresponding linear operator on $\mathrm{im} ( \rho_{\ul{J}} )$ by
\begin{align}\label{phi_defn}
    \iota_{\ul{I}}^{\ul{J}} ( X ) = s_{\ul{J}} \Pi_{\ul{I} , \ul{J}}^{-1} \left( X \otimes \mathbb{I}_{\ul{J} \setminus \ul{I}} \right) \Pi_{\ul{I} , \ul{J}} s_{\ul{J}} \, .
\end{align}
Here we write $\mathbb{I}_{\ul{J} \setminus \ul{I}}$ for the tensor product of identity operators on all subsystems in $\ul{J} \setminus \ul{I}$. In words, the inclusion (\ref{phi_defn}) first re-shuffles tensor product factors to put all sites in $\ul{J} \setminus \ul{I}$ ``at the end'', then tensors $X$ with identity operators on each of these missing sites, and finally re-shuffles tensor product factors to return them to increasing order.

An example might make the construction (\ref{phi_defn}) more transparent. Consider a tripartite Hilbert space $\mathcal{H}_{ABC} = \mathcal{H}_A \otimes \mathcal{H}_B \otimes \mathcal{H}_C$. First suppose that $\ul{I} = ( A, B )$ and $\ul{J} = ( A, B, C )$, and let $X = \mathcal{O}_A \otimes \mathcal{O}_B$ be a linear operator on $\mathrm{im} ( \rho_{AB} )$. In this case, the inclusion (\ref{phi_defn}) is trivial: one simply has
\begin{align}
    \iota_{\ul{I}}^{\ul{J}} ( X ) = \mathcal{O}_A \otimes \mathcal{O}_B \otimes \mathbb{I}_C \big\vert_{\rho_{ABC}} \, ,
\end{align}
which merely tensors with the identity operator on $\mathcal{H}_C$. 

However, now suppose that $\ul{I} = ( A, C )$ and $\ul{J} = ( A, B, C )$ with $X = \mathcal{O}_A \otimes \mathcal{O}_C$. Unfortunately, the tensor product
\begin{align}
    X \otimes \mathbb{I}_B = \mathcal{O}_A \otimes \mathcal{O}_C \otimes \mathbb{I}_B
\end{align}
is a linear operator on $\mathcal{H}_A \otimes \mathcal{H}_C \otimes \mathcal{H}_B$, which has the local Hilbert spaces in the ``wrong order''. The role of the re-shuffling operators $\Pi_{\ul{I} , \ul{J}}$ and $\Pi_{\ul{I} , \ul{J}}^{-1}$ in (\ref{phi_defn}) is to interchange subsystems $C$ and $B$ and place the subsystems in the correct order:
\begin{align}\label{AC_reshuffle_example}
    \iota_{\ul{I}}^{\ul{J}} ( X ) = \mathcal{O}_A \otimes \mathbb{I}_B \otimes \mathcal{O}_C \big\vert_{\rho_{ABC}} \, .
\end{align}
Similarly, for a linear operator $X \in \mathrm{End} ( \mathrm{im} ( \rho_{AC} ) )$ which does not take the factorized form $\mathcal{O}_A \otimes \mathcal{O}_C$ but is rather a linear combination of such operators, the action of $\iota_{\ul{I}}^{\ul{J}} ( X )$ extends (\ref{AC_reshuffle_example}) by linearity.

We are now prepared to define the coboundary operator. Let $\omega \in \Omega^k ( \rho_{\ul{1} \ldots \ul{n}} )$, so that $\omega = ( \omega_{\ul{I}} )$ where $\ul{I}$ runs over all increasing multi-indices of length $k > 0$. For a multi-index $\ul{J} \subset \{ \ul{1} , \ldots, \ul{n} \}$ of length $k+1$, again with elements listed in increasing order
\begin{align}
    \ul{J} = \{ \ul{j}_0 < \ldots < \ul{j}_{k} \} \, ,
\end{align}
we define $d^k \omega$ component-wise by the formula
\begin{align}\label{d_defn}
    \left( d^k \omega \right)_{\ul{J}} = k \sum_{\ell = 0}^{k} ( - 1 )^{\ell} \, \iota_{\ul{J} \setminus \{ \ul{j}_\ell \} }^{\ul{J}} ( \omega_{\ul{J} \setminus \{ \ul{j}_\ell \} } ) \, .
\end{align}
Thus the object $d \omega$ is a tuple of linear operators acting on images of reduced density matrices on $(k+1)$ sites, where each such operator is a linear combination of entries of $\omega$ that have been ``padded with the identity'' on one missing site, each dressed with a sign $\pm 1$ which is correlated with the ordering of the missing site.\footnote{The definition (\ref{d_defn}) is mathematically equivalent to the construction introduced in \cite{Ferko:2024swt}, which used a ``signed tensor product operator'' $\widehat{\otimes}$. Here we have chosen to present the definition differently to emphasize the connection between entanglement cohomology and simplicial cohomology, which we will mention shortly.} The overall factor of $k$ in (\ref{d_defn}), which is unconventional in simplicial cohomology, is introduced for later convenience; in particular, it will ensure that the coboundary satisfies the Leibniz rule with respect to the wedge product to be introduced shortly. However, since this factor is invertible, it does not affect the kernels or images of $d^k$, and thus does not change its cohomology.

Some examples are again in order. For a bipartite Hilbert space $\mathcal{H}_{AB} = \mathcal{H}_A \otimes \mathcal{H}_B$, and an entanglement $1$-form
\begin{align}
    \omega = ( \mathcal{O}_A , \mathcal{O}_B ) \, ,
\end{align}
the coboundary operator acts as
\begin{align}
    d^1 \omega = s_{AB} \left( \mathbb{I}_A \otimes \mathcal{O}_B - \mathcal{O}_A \otimes \mathbb{I}_B \right) s_{AB} \, .
\end{align}
In the case of a tripartite Hilbert space $\mathcal{H}_{ABC} = \mathcal{H}_A \otimes \mathcal{H}_B \otimes \mathcal{H}_C$, an entanglement $1$-form
\begin{align}
    \omega^{(1)} = ( \mathcal{O}_A , \mathcal{O}_B , \mathcal{O}_C ) \, ,
\end{align}
is similarly acted upon by the differential as
\begin{align}
    \hspace{-5pt} d^1 \omega^{(1)} = \left(\left( \mathbb{I}_A \otimes \mathcal{O}_B - \mathcal{O}_A \otimes \mathbb{I}_B \right) \big\vert_{\rho_{AB}} , \left( \mathbb{I}_A \otimes \mathcal{O}_C - \mathcal{O}_A \otimes \mathbb{I}_C \right) \big\vert_{\rho_{AC}} , \left( \mathbb{I}_B \otimes \mathcal{O}_C - \mathcal{O}_B \otimes \mathbb{I}_C \right) \big\vert_{\rho_{BC}} \right) \, .
\end{align}
For an entanglement $2$-form
\begin{align}
    \omega^{(2)} = ( \mathcal{O}_{AB} , \mathcal{O}_{AC} , \mathcal{O}_{BC} ) \, ,
\end{align}
the expression for $d^2 \omega^{(2)}$ requires the action of the permutation operator in order to arrange the tensor product factors in increasing order,
\begin{align}
    d^2 \omega^{(2)} = 2 \left( \mathbb{I}_A \otimes \mathcal{O}_{BC} - \Pi_{AC, ABC}^{-1} \left( \mathcal{O}_{AC} \otimes \mathbb{I}_B \right) \Pi_{AC, ABC} + \mathcal{O}_{AB} \otimes \mathbb{I}_C \right) \big\vert_{\rho_{ABC}} \, ,
\end{align}
where $\Pi_{AC, ABC}^{-1} \left( \mathcal{O}_{AC} \otimes \mathbb{I}_B \right) \Pi_{AC, ABC}$ produces a linear operator acting on $\mathcal{H}_A \otimes \mathcal{H}_B \otimes \mathcal{H}_C$.

We will sometimes suppress the index on the coboundary operator, writing $d$ instead of $d^k$. As in the setting of ordinary differential forms on manifolds, we say that an entanglement $k$-form $\omega$ is \emph{closed} if $d \omega = 0$, and we say that $\omega$ is \emph{exact} if $\omega = d \eta$ for some entanglement $(k-1)$-form $\eta$. It was shown in \cite{Mainiero:2019enr} that these differentials obey
\begin{align}
    d^{k+1} \circ d^k = 0
\end{align}
for all $k$, due to a property of the projectors $s_{\ul{I}}$ called \emph{compatibility of supports}. One may therefore define the associated cohomology groups
\begin{align}
    H^k ( \rho_{\ul{1} \ldots \ul{n}} ) = \frac{\ker ( d^k )}{\mathrm{im} ( d^{k-1} )} \, .
\end{align}
The dimensions of these cohomology groups, viewed as complex vector spaces, are denoted
\begin{align}
    b_k ( \rho_{\ul{1} \ldots \ul{n}} ) = \dim \left( H^k ( \rho_{\ul{1} \ldots \ul{n}} ) \right) \, ,
\end{align}
and called \emph{Betti numbers}.

The generating function for these dimensions is referred to as the \emph{Poincar\'e polynomial},
\begin{align}
    P ( \rho_{\ul{1} \ldots \ul{n}} ; y ) = \sum_{k=1}^{n-1} b_k ( \rho_{\ul{1} \ldots \ul{n}} ) y^{k-1} \, .
\end{align}
We occasionally express the Betti numbers and Poincar\'e polynomials as functions of the pure state which defines the density matrix, writing $P ( \ket{\psi_{\ul{1} \ldots \ul{n}}} ; y )$ rather than $P ( \rho_{\ul{1} \ldots \ul{n}} ; y )$.

Mathematically, the definition (\ref{d_defn}) can be viewed as the usual simplicial coboundary on the $(n-1)$-simplex with vertex set $\{ \ul{1} , \ldots , \ul{n} \}$, but with values in a certain coefficient diagram. Let $\mathcal{C}$ be the category whose objects are subsets $\ul{I} \subset \{ \ul{1} , \ldots , \ul{n} \}$, i.e. faces of the $(n-1)$-simplex $\Upsilon^{n-1}$, and whose morphisms are inclusions $\ul{I} \hookrightarrow \ul{J}$. Define a functor
\begin{align}
    F_{\rho_{\ul{1} \ldots \ul{n}}} : \mathcal{C} \to \mathbf{Vect} \, ,
\end{align}
where $\mathbf{Vect}$ is the category of vector spaces, which acts on objects as
\begin{align}
    F_{\rho_{\ul{1} \ldots \ul{n}}} ( \ul{I} ) = \mathrm{End} ( \mathrm{im} ( \rho_{\ul{I}} ) ) \, ,
\end{align}
and on morphisms as
\begin{align}
    F_{\rho_{\ul{1} \ldots \ul{n}}} ( \ul{I} \hookrightarrow \ul{J} ) = \iota_{\ul{I}}^{\ul{J}} \, .
\end{align}
Then one can interpret the spaces of entanglement $k$-forms for $1 \leq k \leq n$ as
\begin{align}
    \Omega^k ( \rho_{\ul{1} \ldots \ul{n}} ) \cong C^{k-1} ( \Upsilon^{n-1} ; F_{\rho_{\ul{1} \ldots \ul{n}}} ) \, ,
\end{align}
where $C^{k-1}$ denotes the usual space of $(k-1)$-cochains on the $(n-1)$-simplex. Of course, the $(n-1)$-simplex is contractible, so the cohomology groups of $\Upsilon^{n-1}$ with \emph{constant} coefficients are trivial in positive degree, but this is no longer true for cohomology with coefficients that take values in arbitrary diagrams. Nonetheless, there is a close connection between entanglement cohomology and the usual (co)homology of the $(n-1)$-simplex, which we will exploit in Section \ref{sec:simplicial_lemma} to prove a lemma which is useful for computing the entanglement cohomologies of generalized GHZ and W states. A review of standard material about simplicial homology and the $(n-1)$-simplex $\Upsilon^{n-1}$ can be found in Appendix \ref{app:simp}.

We assemble the spaces of entanglement $k$-forms into a cochain complex,
\begin{align}
    0 \to \mathbb{C} \xrightarrow{d^0} \Omega^1 ( \rho_{\ul{1} \ldots \ul{n}} )  \xrightarrow{d^1} \Omega^2 ( \rho_{\ul{1} \ldots \ul{n}} )  \xrightarrow{d^2} \ldots \xrightarrow{d^{n-1}} \Omega^{n} ( \rho_{\ul{1} \ldots \ul{n}} )  \xrightarrow{d^{n}} 0 \, ,
\end{align}
where the $0$-th coboundary operator is defined to act on scalars $\lambda \in \mathbb{C}$ as
\begin{align}\label{d0_defn}
    d^0 \lambda = \bigtimes_{\ul{i} = \ul{1}}^{\ul{n}} \lambda s_{\ul{i}}  = ( \lambda s_{\ul{1}} , \ldots , \lambda s_{\ul{n}} ) \, .
\end{align}
As we alluded to in the introduction, the structure of entanglement $k$-forms shares several properties with that of differential forms defined on a manifold with a metric. For instance, by virtue of the Schmidt decomposition, one can define a version of the Hodge star operation on entanglement forms. For any multi-index $\ul{I}$, one can express the parent pure state (\ref{pure_state_defn}) defining our entanglement complex as
\begin{align}\label{schmidt}
    \ket{\psi_{\ul{1} \ldots \ul{n}}} = \sum_{\alpha = 1}^{S} \lambda_\alpha \ket{\alpha_{\ul{I}}} \otimes \big\vert \alpha_{\ul{I}^C} \big\rangle \, ,
\end{align}
up to the action of a permutation operation which re-shuffles the subsystems in $\ul{I} \cup \ul{I}^C$ into the order $\{ \ul{1} , \ldots , \ul{n} \}$, which we omit for simplicity. Here $S$ is an integer called the \emph{Schmidt rank}, not to be confused with the symmetric group $S_n$ or a support projection $s_{\ul{i}}$, and the quantities $\lambda_\alpha$ are numbers known as \emph{Schmidt coefficients}. The states $\ket{\alpha_{\ul{I}}}$, $\big\vert \alpha_{\ul{I}^C} \big\rangle$ belong to $\mathcal{H}_{\ul{I}}$ and $\mathcal{H}_{\ul{I}^C}$ and span the images of $\rho_{\ul{I}}$ and $\rho_{\ul{I}^C}$, respectively. Therefore, given an entanglement $k$-form $\omega$ with a component $\omega_{\ul{I}} \in \mathrm{End} ( \mathrm{im} ( \rho_{\ul{I}} ) )$, we can express the matrix elements of this operator in the Schmidt basis as
\begin{align}
    \omega_{\ul{I}} = \sum_{\alpha, \beta = 1}^{S} \left( \omega_{\ul{I}} \right)_{\alpha \beta} \ket{\alpha_{\ul{I}}} \bra{\beta_{\ul{I}}} \, .
\end{align}
We define the component of an entanglement $(n-k)$ form $\ast \omega$ which is associated with the complement $\ul{I}^C$ by the expression
\begin{align}\label{hodge_defn}
    \left( \ast \omega \right)_{\ul{I}^C} = ( - 1 )^\sigma \sum_{\alpha, \beta = 1}^{S} \left( \omega_{\ul{I}}^\ast \right)_{\alpha \beta} \big\vert \alpha_{\ul{I}^C} \big\rangle \big\langle \beta_{\ul{I}^C} \big\vert  \, ,
\end{align}
where $\sigma \in S_{\ul{n}}$ is the permutation of $\{ \ul{1} , \ldots , \ul{n} \}$ given by $\ul{I} \cup \ul{I}^C$, and the collection of all such components as $\ul{I}$ varies over length-$k$ multi-indices defines $\ast \omega$. On the right side of (\ref{hodge_defn}), the symbol $\left( \omega_{\ul{I}}^\ast \right)_{\alpha \beta}$ denotes the complex conjugate of $\left( \omega_{\ul{I}} \right)_{\alpha \beta}$. Like the Hodge star acting on differential forms, this operation on $k$-forms is an involution up to a sign:
\begin{align}
    \ast \ast = ( - 1 )^{k ( n - k )} \, .
\end{align}
Given an entanglement $k$-form $\omega$ and an entanglement $p$-form $\eta$, with $k + p \leq n$, one can also define their wedge product $\omega \wedge \eta$. For any length $(k+p)$ multi-index $\ul{K}$, the corresponding component of this wedge product is
\begin{align}\label{wedge_definition}
    \left( \omega \wedge \eta \right)_{\ul{K}} = \sum_{\ul{I} \sqcup \ul{J} = \ul{K}} ( - 1 )^\sigma \left( s_{\ul{K}} \Pi_{\ul{I} , \ul{K}}^{-1} \left( \omega_{\ul{I}} \otimes \eta_{\ul{J}} \right) \Pi_{\ul{I} , \ul{K}} s_{\ul{K}}  \right) \, ,
\end{align}
where the sum runs over all pairs of multi-indices $\ul{I}$, $\ul{J}$ whose disjoint union is $\ul{K}$, and for each term the factor $(-1)^\sigma$ is the sign of the permutation mapping $\ul{I} \sqcup \ul{J}$ to $\ul{K}$. 

For instance, on a bipartite Hilbert space $\mathcal{H}_{AB}$ with entanglement $1$-forms
\begin{align}
    \omega = ( \omega_A, \omega_B ) \, , \qquad \eta = ( \eta_A, \eta_B ) \, ,
\end{align}
this wedge product acts as the usual wedge product on differential one-forms,
\begin{align}
    \omega \wedge \eta = ( \omega_A \otimes \eta_B - \eta_A \otimes \omega_B ) \big\vert_{\rho_{AB}} \, .
\end{align}
Combining this definition of the wedge product with the Hodge star operation, one can define an inner product on $\Omega^k ( \rho_{\ul{1} \ldots \ul{n}} )$ by
\begin{align}
    \langle \omega , \eta \rangle = \tr \left( \left( \omega \wedge \ast \eta \right) \big\vert_{\rho_{\ul{1} \ldots \ul{n}}} \right) \, ,
\end{align}
where the trace is taken in the total Hilbert space $\mathcal{H}_{\ul{1} \ldots \ul{n}}$, or more precisely in its one-dimensional subspace $\mathrm{im} ( \rho_{\ul{1} \ldots \ul{n}} )$.

The factor of $k$ which was included in the definition (\ref{d_defn}) ensures that this notion of wedge product satisfies the Leibniz formula. If $\omega$ is an entanglement $p$-form, one has
\begin{align}
    d \left( \omega \wedge \eta \right) = \left( d \omega \right) \wedge \eta + ( - 1 )^p \omega \wedge \left( d \eta \right) \, .
\end{align}
Properties of the various operations on entanglement $k$-forms are discussed in \cite{Ferko:2024swt}, to which we refer the reader for further details. We will review only one additional construction here. Using the inner product, we define the adjoint $\delta$ of the coboundary operator,
\begin{align}\label{adjoint_defn}
    \langle \omega , d \eta \rangle = \langle \delta \omega, \eta \rangle \, ,
\end{align}
and the Laplacian operator given in equation (\ref{lap_defn}), which we repeat:
\begin{align}\label{lap_defn_later}
    \Delta = d \delta + \delta d \, .
\end{align}

\section{Proofs of Conjectures for GHZ and W States}\label{sec:proofs}

It has been known for decades that a collection of three qubits can be entangled in two inequivalent ways \cite{Dur:2000zz}, exemplified by the GHZ state
\begin{align}\label{GHZ3}
    \ket{ \mathrm{GHZ} } = \frac{1}{\sqrt{2}} \left( \ket{000} + \ket{111} \right) \, ,
\end{align}
and the W state,
\begin{align}\label{W3}
    \ket{ \mathrm{W} } = \frac{1}{\sqrt{3}} \left( \ket{001} + \ket{010} + \ket{100} \right) \, .
\end{align}
The states (\ref{GHZ3}) and (\ref{W3}) exhibit qualitatively different physical features. For instance, the GHZ state is fragile to particle loss: upon tracing out a single qubit, the state (\ref{GHZ3}) reduces to a classical mixture of product states with no bipartite entanglement between the remaining two qubits. In contrast, tracing out one qubit in the W state (\ref{W3}) gives a reduced system in which the two remaining qubits are still entangled.

One would expect that this qualitative difference is encoded in some feature of the entanglement $k$-forms associated with the states (\ref{GHZ3}) and (\ref{W3}). Indeed, the distinction is already visible from the dimensions of their cohomologies, i.e. their Poincar\'e polynomials:
\begin{align}
    P \left( \ket{ \mathrm{W} } ; y \right) = 3 + 3 y \, , \qquad P \left( \ket{ \mathrm{GHZ} } ; y \right) = 7 + 7 y \, .
\end{align}
Because the dimensions of entanglement cohomologies are invariant under local unitary transformations, this result gives another way of understanding the LU inequivalence of the tripartite GHZ and W states.

It is natural to wonder about the mathematical structure of the cohomologies for generalizations of the states (\ref{GHZ3}) and (\ref{W3}) to larger numbers of subsystems and different choices of local Hilbert space dimensions. Fix an $n$-partite Hilbert space of the general form (\ref{general_hilbert_space}), with dimensions described by the vector $\vec{d} = ( d_{\ul{1}} , \ldots, d_{\ul{n}} )$ where $d_{\ul{i}} \geq 2$ for all $\ul{i}$, and define the generalized GHZ state
\begin{align}\label{generalized_ghz_defn}
    \ket{ \mathrm{GHZ}_{n, \vec{d} } } = \frac{1}{\sqrt{2}} \left( \ket{0}^{\otimes n} + \ket{1}^{\otimes n} \right) \, ,
\end{align}
along with the generalized W state,
\begin{align}\label{generalized_W}
    \ket{ \text{W}_{n, \vec{d} } } = \frac{1}{\sqrt{n}} \left( \sum_{i = 1}^{n} \ket{0}^{\otimes ( i - 1 ) } \otimes \ket{1} \otimes \ket{0}^{\otimes (n - i ) } \right) \, .
\end{align}
This notation (\ref{generalized_W}) means that, in the $i$-th term of the sum, the $i$-th subsystem occurs with a state $\ket{1}$ and all other subsystems appear with a state $\ket{0}$. Here we use the obvious shorthand ${}^{\otimes k}$ to denote a $k$-fold tensor product,
\begin{align}
    \ket{0}^{\otimes k} = \underbrace{ \ket{0} \otimes \ldots \otimes \ket{0} }_{k \text{ times}} \, ,
\end{align}
and likewise for $\ket{1}^{\otimes k}$. To simplify notation, it is convenient to define
\begin{align}
    \ket{e_i} = \ket{0}^{\otimes ( i - 1 ) } \otimes \ket{1} \otimes \ket{0}^{\otimes ( n - i )} = \big| \underbrace{0 \ldots 0}_{i-1 \text{ times } }  1 \underbrace{ 0 \ldots 0}_{n-i \text{ times} } \big\rangle \, ,
\end{align}
which has a $1$ in slot $i$ and $0$ elsewhere. In this notation,
\begin{align}\label{nice_W_expr}
    \ket{ \text{W}_{n , \vec{d} } } = \frac{1}{\sqrt{n}} \sum_{i=1}^{n} \ket{e_i} \, .
\end{align}
When $n = 2$, both the GHZ and W states reduce to a Bell pair, so we will restrict attention to $n \geq 3$ in what follows. Using numerical code to compute the dimensions of entanglement cohomologies for the first few non-trivial GHZ and W states, one finds the Poincar\'e polynomials listed in Table \ref{tab:ghz-w-poincare}.

\begin{table}[h]
  \centering
\begin{tabular}{ |p{0.5cm}||p{9cm}|p{2cm}|  }
 \hline
 \multicolumn{3}{|c|}{Numerical Poincar\'e Polynomials of GHZ and W States} \\
 \hline
 $n$ & \multicolumn{1}{c|}{$P ( \big| \mathrm{GHZ}_{n, \vec{d}} \big\rangle ; y )$}
     & \multicolumn{1}{c|}{$P ( \big| \mathrm{W}_{n, \vec{d}} \big\rangle ; y )$} \\
 \hline
 $3$  & $7 + 7 y$    & $3 + 3 y$\\
 $4$ &  $9 + 12 y + 9 y^2$ & $3 + 3 y^2$ \\
 $5$ & $11 + 20 y + 20 y^2 + 11 y^3$ & $3 + 3 y^3$ \\
 $6$ & $13 + 30 y + 40 y^2 + 30 y^3 + 13 y^4$ & $3 + 3 y^4$ \\
 $7$ &  $15 + 42 y + 70 y^2 + 70 y^3 + 42 y^4 + 15 y^5$  & $3 + 3 y^5$ \\
 $8$ & $17 + 56 y + 112 y^2 + 140 y^3 + 112 y^4 + 56 y^5 + 17 y^6$  & $3 + 3 y^6$  \\
 $9$ & $19 + 72 y + 168 y^2 + 252 y^3 + 252 y^4 + 168 y^5 + 72 y^6 + 19 y^7$  & $3 + 3 y^7$\\
 \hline
\end{tabular}
  \caption{Experimental values of the dimensions of cohomologies of generalized GHZ and W states are computed for various $n$ using a Python implementation. We create pure states as QuTiP \cite{Johansson:2011jer,Johansson:2012qtx} objects, form their density matrices, and compute all reduced density matrices along with their ranks and support projections. We then enumerate an explicit basis of $\mathrm{End} ( \mathrm{im} ( \rho_{\ul{I}} ) )$ for each multi-index $\ul{I}$, represented as matrices, and use this to form a basis for each space $\Omega^k ( \rho )$ of entanglement $k$-forms. Finally, we form matrix representations of each coboundary operator $d^k$ defined in (\ref{d_defn}) by acting with $d^k$ on each basis element of $\Omega^k ( \rho )$ and expressing the result in a basis for $\Omega^{k+1} ( \rho )$. Computing the images and kernels of the matrix representations of each such operator $d^k$ then allows us to find the dimensions of cohomologies, $\dim ( H^k ) = \dim ( \mathrm{ker} ( d^k ) ) - \dim ( \mathrm{im} ( d^{k-1} ) )$.}
  \label{tab:ghz-w-poincare}
\end{table}

In particular, one finds the same Poincar\'e polynomials at fixed $n$ regardless of the specific values of the local Hilbert space dimensions $d_{\ul{i}}$ appearing in the vector $\vec{d}$, as one might expect since the definitions of the generalized GHZ and W states only involve the computational basis elements $\ket{0}$ and $\ket{1}$, so any other basis vectors are simply spectators.

In view of this fact, to ease notation, we will omit the symbol $\vec{d}$ and define the shorthand
\begin{align}
    P^{\mathrm{GHZ}}_n ( y ) = P \left( \big| \mathrm{GHZ}_{n, \vec{d}} \big\rangle ; y \right) \, , \qquad P^{\mathrm{W}}_n ( y ) = P \left( \big| \mathrm{W}_{n, \vec{d}} \big\rangle ; y \right) \, .
\end{align}
Based on this numerical evidence, in \cite{Mainiero:2019enr} it was conjectured that the Poincar\'e polynomials for all generalized GHZ and W states take simple closed-form expressions.
\begin{conj}\label{GHZ_conj}
    For all $n \geq 3$ and dimension vectors $\vec{d}$,
    \begin{align}\label{GHZ_polynomial}
        P^{\mathrm{GHZ}}_n ( y ) &= 1 + y^{n-2} + \frac{2}{y} \left( \left( 1 + y \right)^n - \left( 1 + y^n \right) \right) \nonumber \\
        &= \left( 2 n + 1 \right) + 2 \sum_{m=2}^{n-2} {n \choose m} y^{m-1} + \left( 2 n + 1 \right) y^{n-2} \, .
    \end{align}
\end{conj}
The coefficients in (\ref{GHZ_polynomial}) are symmetric -- as is required by Hodge duality -- and grow linearly in $n$ at the ends but as binomial coefficients in the interior. As was already pointed out in \cite{Mainiero:2019enr}, these coefficients follow the structure of a modified Pascal's triangle.
\begin{conj}\label{W_conj}
    For all $n \geq 3$ and dimension vectors $\vec{d}$,
    \begin{align}
        P^{\mathrm{W}}_n ( y ) = 3 + 3 y^{n-2} \, .
    \end{align}
\end{conj}
That is, generalized W states are believed to have only two non-vanishing Betti numbers, $b_1 = b_{n-1} = 3$, independent of $n$ and of the local dimensions. Again, this is compatible with Hodge duality, which requires that the top and bottom cohomologies are isomorphic.

There is strong numerical evidence for Conjectures \ref{GHZ_conj} and \ref{W_conj}, but to the best of our knowledge, no explicit proof has yet been presented. The goal of this section is to rigorously establish these two results. The proof of each of these conjectures splits into two parts. The first part concerns the details of the structure of the states under consideration, and in particular the images of their reduced density matrices and the action of the operators $d^k$ on tuples of operators acting on each space $\mathrm{End} ( \mathrm{im} ( \rho_{\ul{I}} ) )$. Once the explicit action of each such $d^k$ has been determined, the second part of the proof is common to both conjectures, and simply involves a counting argument to determine the dimensions of the kernels and images of the operators $d^k$. Because this second step of the proof is shared in both conjectures, we will present it first as a lemma in Section \ref{sec:simplicial_lemma}. Although this result is elementary, it clarifies how the structure of entanglement cohomology -- at least for the special cases of generalized GHZ and W states -- is related to ordinary simplicial homology, and in particular the contractibility of the $n$-simplex. After establishing this basic lemma, we apply it in Sections \ref{sec:ghz_proof} and \ref{sec:w_proof} to prove Conjectures \ref{GHZ_conj} and \ref{W_conj}, respectively.

\subsection{Simplicial Lemma}\label{sec:simplicial_lemma}

As we mentioned around equation (\ref{d_defn}), entanglement cohomology can be viewed as a certain modification of the simplicial cohomology of the $(n-1)$-simplex with coefficients involving endomorphisms of appropriate reduced density matrices. In our proofs of Conjectures \ref{GHZ_conj} and \ref{W_conj}, we will ``trivialize'' these coefficients by working in explicit bases, which reduces the computation of the cohomologies to repeated application of standard facts about simplicial cohomology. The primary result which we will use is the following lemma.

\begin{lemma}\label{simplex_lemma}
    Fix positive integers $n$ and $k$ with $1 \leq k \leq n-1$, and let $V_n = \{ 1, \ldots, n \}$. For each $k$-element subset $I \subset V_n$, define a (real or complex) variable $x_I$. For convenience, we will always assume that subsets are sorted as $I = \{ i_1 , \ldots, i_k \}$ with $i_1 < \ldots < i_k$. Before imposing any constraints, the (real or complex) dimension of the space of all $x_I$ is ${n \choose k}$.
    
    Now suppose that, for every $(k+1)$-element subset $J = \{ j_0 , \ldots, j_k \} \subset V_n$, again assuming the sorting condition $j_0 < \ldots < j_k$, we impose the alternating sum constraint
    \begin{align}\label{alternating_constraint}
        \sum_{r = 0}^{k} ( - 1 )^r x_{J \setminus \{ j_r \} } = 0 \, .
    \end{align}
    Then the solution space of (\ref{alternating_constraint}) has (real or complex) dimension ${n-1 \choose k-1}$.
\end{lemma}

Lemma \ref{simplex_lemma} is essentially the statement that the $(n-1)$-simplex is contractible. An intuitive way to think of the proof is that the collection of variables $( x_I )$ is uniquely determined by its values on $k$-element subsets that contain a single fixed element of $V_n$, which we may take to be $1 \in V_n$ without loss of generality. In particular, suppose that we know the values of the variables $( x_I )$ on all subsets of $V_n$ containing $1$, and we wish to determine the value of $x_I$ on some other subset $I = \{ i_1, \ldots, i_k \}$ which does not contain $1$. Form the subset $J = \{ 1 \} \cup I$, which is a $(k+1)$-element subset of $V_n$, and thus we can solve the constraint (\ref{alternating_constraint}):
\begin{align}\label{solved_constraint}
    x_{i_1 \ldots i_k} = \sum_{r = 1}^{k} ( - 1 )^{r-1} x_{1 i_1 \ldots \widehat{i}_r \ldots i_k} \, ,
\end{align}
where we write a hatted index $\widehat{i}_r$ to indicate that it is omitted from the sum. Every term on the right side of (\ref{solved_constraint}) involves a multi-index containing the entry $1$, so it is known by assumption. Furthermore, there is one such constraint (\ref{solved_constraint}) for each subset $I \subset V_n$ which does not contain $1$, so we have the same number of unknown variables as equations. Thus we can solve for all of the variables $( x_I )$ given only the values of $x_I$ on subsets $I$ which contain a single fixed index like $1$, and the number of such values is ${n-1 \choose k-1}$, which establishes the claim.

Although the above argument gives the idea of why Lemma \ref{simplex_lemma} is true, we find it instructive to include a more formal proof, which also explains why this result is essentially an immediate consequence of contractibility of the simplex.

\begin{proof}
    Let $\Upsilon^{n-1}$ be the $(n - 1)$ simplex on $n$ vertices. A $(k-1)$-cochain $x \in C^{k-1} ( \Upsilon^{n-1} ) $ on $\Upsilon^{n-1}$ is a function that assigns a scalar to each $(k-1)$-simplex, i.e. to each $k$-element subset $I \subset V_n$, which is precisely a collection of ${n \choose k}$ variables $x \equiv ( x_I )$.

    We choose the standard orientation convention where the oriented $k$-face with ordered vertices $j_0 < \ldots < j_k$ has boundary
    \begin{align}
        \partial [ j_0 \ldots j_k ] = \sum_{r=0}^{k} ( - 1 )^r [ j_0 \ldots \widehat{j}_r \ldots j_k ] \, ,
    \end{align}
    where again a hat denotes omission in the sum.

    The corresponding coboundary operator on cochains, $\delta : C^{k-1} \to C^k$, is\footnote{We trust that this operator will not be confused with the codifferential acting on entanglement forms, defined in equation (\ref{adjoint_defn}), since the difference will be clear from context.}
    \begin{align}
        ( \delta x ) \left( [ j_0 \ldots j_k ] \right) &= x \left( \partial [ j_0 \ldots j_k ] \right) = \sum_{r=0}^{k} ( - 1 )^r x_{j_0 \ldots \widehat{j}_r \ldots j_k} \, .
    \end{align}
    In this language, the alternating sum constraint (\ref{alternating_constraint}) is simply
    \begin{align}\label{x_cocycle}
        \delta x = 0 \, ,
    \end{align}
    which means that $x$ is a cocycle. The solution space is therefore
    \begin{align}
        Z^{k-1} ( \Upsilon^{n-1} ) = \ker \left( \delta : C^{k-1} \to C^k \right) \, .
    \end{align}
    We wish to show that
    \begin{align}
        \dim ( Z^{k-1} ) = {n-1 \choose k-1} \, .
    \end{align}
    To do this, we use the familiar fact from algebraic topology that $\Upsilon^{n-1}$ is contractible, so
    \begin{align}
        H^p ( \Upsilon^{n-1} ) = 0 
    \end{align}
    for all $p \geq 1$, while $\dim \left( H^0 ( \Upsilon^{n-1} ) \right) = 1$. Since $H^p = \frac{Z^p}{B^p}$, where $B^p = \mathrm{im} ( \delta : C^{p-1} \to C^p )$, this means that $Z^p = B^p$ for $p \geq 1$ and $\dim ( Z^0 ) = 1$.

    Letting $a_k = \dim ( Z^{k-1} )$, for $k \geq 2$ we have
    \begin{align}
        a_k = \dim ( B^{k-1} ) = \mathrm{rank} ( \delta : C^{k-2} \to C^{k-1} ) \, ,
    \end{align}
    but by the rank-nullity theorem,
    \begin{align}
        \mathrm{rank} ( \delta ) = \dim ( C^{k-2} ) - \dim ( \ker ( \delta ) ) = {n \choose k - 1} - a_{k-1} \, .
    \end{align}
    This means that we have the base case and recursion relation
    \begin{align}
        a_1 = 1 \, , \qquad a_k = {n \choose k - 1} - a_{k-1} \text{ for } k \geq 2 \, .
    \end{align}
    The  solution to this recursion is $a_k = { n - 1 \choose k - 1 }$. The base case is clear: $a_1 = {n - 1 \choose 0} = 1$. And assuming $a_{k-1} = { n - 1 \choose k- 2}$, we find
    \begin{align}
        a_k = { n \choose k - 1 } - { n - 1 \choose k - 2} = { n - 1 \choose k - 1 } \, ,
    \end{align}
    by Pascal's identity. This establishes the lemma for all $n$ and $1 \leq k \leq n - 1$.
\end{proof}

\subsection{GHZ States}\label{sec:ghz_proof}

We will now compute the entanglement cohomologies for the generalized GHZ states (\ref{generalized_ghz_defn}). 

\begin{proof}
As we mentioned above, the specific vector $\vec{d}$ of local dimensions will not play any role in our discussion, so we will suppress it and write the associated density matrix as
\begin{align}
    \rho^{\text{GHZ}}_{\ul{1} \ldots \ul{n}} = \frac{1}{2} \left( \ket{ 0^{\otimes n} } \bra{ 0^{\otimes n} } + \ket{0^{\otimes n}} \bra{1^{\otimes n}} + \ket{1^{\otimes n}} \bra{0^{\otimes n}} + \ket{1^{\otimes n}} \bra{1^{\otimes n}} \right) \, .
\end{align}
To form the spaces $\mathrm{End} ( \mathrm{im} ( \rho_{\ul{I}} ) )$ involving the reduced density matrices, we will need to evaluate various partial traces. Let $\ul{I}^C$ be a length-$k$ multi-index of subsystems, for $2 \leq k \leq n - 2$ which we wish to trace out, retaining the subsystems associated with the length-$m$ multi-index $\ul{I}$ where $m = n - k$. The partial traces are trivial to compute,
\begin{align}\label{ghz_partial_traces}
    \tr_{\ul{I}^C} \left( \ket{ 0^{\otimes n} } \bra{ 0^{\otimes n}} \right) &= \ket{ 0^{\otimes m} } \bra{ 0^{\otimes m}} \, , \nonumber \\
    \tr_{\ul{I}^C} \left( \ket{ 0^{\otimes n} } \bra{ 1^{\otimes n}} \right) &= 0 \, , \nonumber \\
    \tr_{\ul{I}^C} \left( \ket{ 1^{\otimes n} } \bra{ 0^{\otimes n}} \right) &= 0 \, , \nonumber \\
    \tr_{\ul{I}^C} \left( \ket{ 1^{\otimes n} } \bra{ 1^{\otimes n}} \right) &= \ket{ 1^{\otimes m} } \bra{ 1^{\otimes m}} \, ,
\end{align}
where we use the notation ${}^{\otimes m}$ to refer to tensor products of states in the $m$ factors of $\ul{I}$. Explicitly, each of the operators in (\ref{ghz_partial_traces}) act on
\begin{align}
    \mathcal{H}_{\ul{I}} = \bigotimes_{\ul{i} \in \ul{I}} \mathcal{H}_{\ul{i}} \, ,
\end{align}
and all of the reduced density matrices take the form
\begin{align}\label{GHZ_reduced_S}
    \rho^{\mathrm{GHZ}}_{\ul{I}} = \frac{1}{2} \left( \ket{ 0^{\otimes m} } \bra{ 0^{\otimes m}} + \ket{ 1^{\otimes m} } \bra{ 1^{\otimes m}} \right) \, ,
\end{align}
so the support projections
\begin{align}
    s^{\mathrm{GHZ}}_{\ul{I}} = \ket{ 0^{\otimes m} } \bra{ 0^{\otimes m}} + \ket{ 1^{\otimes m} } \bra{ 1^{\otimes m}}
\end{align}
are rank-$2$. Thus any operator $\mathcal{O}_{\ul{I}} \in \mathrm{End} ( \mathrm{im} ( \rho^{\mathrm{GHZ}}_{\ul{I}} ) )$ is determined by four matrix elements
\begin{align}\label{GHZ_O_defn}
    \mathcal{O}_{\ul{I}} = a_{\ul{I}} \ket{ 0^{\otimes m} } \bra{ 0^{\otimes m}} + b_{\ul{I}} \ket{ 0^{\otimes m} } \bra{ 1^{\otimes m}} + c_{\ul{I}} \ket{ 1^{\otimes m} } \bra{ 0^{\otimes m}} + d_{\ul{I}} \ket{ 1^{\otimes m} } \bra{ 1^{\otimes m}} 
\end{align}
in this basis. 

Now fix a subsystem $\ul{j} \notin \ul{I}$ and consider the length $(m+1)$ multi-index $\ul{J} = \{ \ul{j} \} \cup \ul{I}$. We wish to compute the matrix elements of terms of the form appearing in the coboundary formula (\ref{d_defn}) which arise from the contributions of $\mathcal{O}_{\ul{I}}$. Each such operator morally takes the form of a tensor product $\mathbb{I}_{\ul{j}} \otimes \mathcal{O}_{\ul{I}}$, up to a sign depending on the permutation and a re-shuffling of tensor product factors. By a direct computation one finds that
\begin{align}\label{GHZ_d_basis_expansion}
    \bra{ 0^{\otimes ( m + 1 ) } } \iota_{\ul{I}}^{\ul{J}} ( \mathcal{O}_{\ul{I}} ) \ket{ 0^{\otimes ( m + 1 ) } } &= a_{\ul{I}} \, , \nonumber \\
    \bra{ 0^{\otimes ( m + 1 ) } } \iota_{\ul{I}}^{\ul{J}} ( \mathcal{O}_{\ul{I}} ) \ket{ 1^{\otimes ( m + 1 ) } } &= 0 \, , \nonumber \\
    \bra{ 1^{\otimes ( m + 1 ) } } \iota_{\ul{I}}^{\ul{J}} ( \mathcal{O}_{\ul{I}} ) \ket{ 0^{\otimes ( m + 1 ) } } &= 0 \, , \nonumber \\
    \bra{ 1^{\otimes ( m + 1 ) } } \iota_{\ul{I}}^{\ul{J}} ( \mathcal{O}_{\ul{I}} ) \ket{ 1^{\otimes ( m + 1 ) } } &= d_{\ul{I}} \, ,
\end{align}
where we use the natural basis for $\mathrm{im} ( \rho_{\ul{J}} )$. 

We may now use this fact to compute $\dim ( \ker ( d^m ) )$. Let $\omega = ( \mathcal{O}_{\ul{I}} )$ be an entanglement $m$-form whose components take the form (\ref{GHZ_O_defn}) for each length-$m$ multi-index $\ul{I}$. All of the components in the sum (\ref{d_defn}) vanish if and only if all four of the matrix elements (\ref{GHZ_d_basis_expansion}) associated with each $( d \omega )_{\ul{J}}$ vanish. We see that every entry $c_{\ul{I}}$ and $b_{\ul{I}}$ is in this kernel, since those matrix elements do not appear in (\ref{GHZ_d_basis_expansion}) and thus are annihilated by the map $d^m$. There are also contributions to the kernel from the solution space arising from the $a_{\ul{I}}$ and $d_{\ul{I}}$ that are annihilated by the coboundary map. From the definition (\ref{d_defn}), we see that the conditions for a combination of $a_{\ul{I}}$ or $d_{\ul{I}}$ to lie in the kernel are
\begin{align}\label{alternating_constraint_ghz}
    \sum_{r = 0}^{m} ( - 1 )^r a_{ \ul{J} \setminus \{ \ul{j}_r \} } = 0 \, , \qquad \sum_{r = 0}^{m} ( - 1 )^r d_{ \ul{J} \setminus \{ \ul{j}_r \} } = 0 \, ,
\end{align}
respectively. By Lemma \ref{simplex_lemma}, the solution space of each of these two constraints is ${n - 1 \choose m - 1}$, so
\begin{align}
    \dim ( \mathrm{ker} ( d^m ) ) = 2 {n - 1 \choose m - 1} + 2 {n \choose m} \, ,
\end{align}
where the first term comes from the solution space of all of the $a_{\ul{I}}$ and $d_{\ul{I}}$ which are mapped to zero, and the second term comes from the $b_{\ul{I}}$ and $c_{\ul{I}}$ that are automatically in the kernel.

The dimension of the $m$-th cohomology group, where again $2 \leq m \leq n - 2$, is then
\begin{align}
    \dim ( H^m ) &= \dim ( \ker ( d^m ) ) - \dim ( \mathrm{im} ( d^{m-1} ) ) \nonumber \\
    &=  \dim ( \ker ( d^m ) ) - \left( \dim ( \Omega^{m-1} ) - \dim ( \ker ( d^{m-1} ) ) \right) \nonumber \\
    &= 2 {n - 1 \choose m - 1} + 2 {n \choose m} - 4 {n \choose m - 1} + 2 {n - 1 \choose m - 2} + 2 {n \choose m - 1} \nonumber \\
    &= 2 {n \choose m} \, ,
\end{align}
where we have used the rank-nullity theorem, the fact that $\dim \left( \Omega^m \right) = 4 {n \choose m}$ as there are ${n \choose m}$ length-$m$ multi-indices and each operator has four independent entries, and Pascal's identity. This confirms the coefficients in the conjectured Poincar\'e polynomial on the second line of (\ref{GHZ_polynomial}), except for the lowest and highest Betti numbers, which we have excluded by taking $2 \leq k \leq n-2$. To complete the proof, we handle these edge cases directly. Since the image of the map $d^0$ defined in equation (\ref{d0_defn}) is clearly one-dimensional,
\begin{align}
    \dim ( H^1 ) &= \dim ( \ker ( d^1 ) ) - \dim ( \mathrm{im} ( d^0 ) ) \nonumber \\
    &= 2 {n - 1 \choose 0} + 2 {n \choose 1} - 1 \nonumber \\
    &= 2 n + 1 \, .
\end{align}
Similarly, as $\dim ( \mathrm{im} ( d^{n-1} ) ) = 1$, we have
\begin{align}
    \dim ( H^{n-1} ) &= \dim ( \ker ( d^{n-1} ) ) - \dim ( \mathrm{im} ( d^{n-2} ) ) \nonumber \\
    &= \dim ( \Omega^{n-1} ) - 1 - \left( \dim ( \Omega^{n-2} ) - \dim ( \ker ( d^{n-2} ) ) \right) \nonumber \\
    &= 4 {n \choose n - 1} - 1 - 4 { n \choose n - 2 } + 2 { n - 1 \choose n - 3} + 2 { n \choose n - 2 } \nonumber \\
    &= 2 n + 1 \, .
\end{align}
These final two dimensions also agree with (\ref{GHZ_polynomial}), establishing Conjecture \ref{GHZ_conj}.
\end{proof}

We see that the proof of Conjecture \ref{GHZ_conj} crucially relied on certain special properties of generalized GHZ states -- namely, that all of their reduced density matrices are rank-$2$, and that the action of the coboundary operator in a natural basis takes a certain consistent structure across entanglement forms of different degrees.

\subsection{W States}\label{sec:w_proof}

Next we apply a similar technique to prove Conjecture \ref{W_conj} regarding the dimensions of cohomologies for generalized W states. As we will see, the reduced density matrices are again all rank-$2$ in this case, but the action of the coboundary operator will now involve all four matrix elements in an appropriate basis.

\begin{proof}

Using the notation (\ref{nice_W_expr}), the density matrix associated to the W state is
\begin{align}
    \rho^{\text{W}}_{\ul{1} \ldots \ul{n}} = \frac{1}{n} \sum_{\ul{i}, \ul{j} = \ul{1}}^{\ul{n}} \ket{e_\ul{i}} \bra{e_\ul{j}} \, ,
\end{align}
where again we suppress the vector $\vec{d}$ of subsystem dimensions, since all states beyond $\ket{0}$ and $\ket{1}$ are spectators.

As before, it is convenient to develop some formulas for reduced density matrices. Suppose that we wish to trace out a length-$k$ multi-index $\ul{I}^C$ of subsystems, again with $2 \leq k \leq n- 2$, while keeping the subsystems labeled by the length $m = n - k$ multi-index $\ul{I}$. In addition to the states $\ket{e_{\ul{i}}} \in \mathcal{H}_{\ul{1} \ldots \ul{n}}$, define analogous states
\begin{align}
    \ket{f_{\ul{r}}} = \ket{0}^{\otimes ( r - 1 )} \otimes \ket{1} \otimes \ket{0}^{\otimes ( m - r )} \in \mathcal{H}_{\ul{I}} \, ,
\end{align}
where we take indices $\ul{r}, \ul{s} = \ul{i}_1 \ldots \ul{i}_m$ to label subsystems in the reduced Hilbert space $\mathcal{H}_{\ul{I}}$ and non-underlined variables $r, s = 1 , \ldots, m$ to label the integers associated with the corresponding tensor product factors.

Now consider a particular term in the partial trace
\begin{align}
    \tr_{\ul{I}^C} \left( \rho^{\text{W}}_{\ul{1} \ldots \ul{n}} \right) = \frac{1}{n} \sum_{\ul{i}, \ul{j} = \ul{1}}^{\ul{n}} \tr_{\ul{I}^C} \left( \ket{e_\ul{i}} \bra{e_\ul{j}} \right) \, .
\end{align}
There are three possibilities giving three sets of contributions:
\begin{enumerate}
    \item Both $\ul{i}$ and $\ul{j}$ belong to $\ul{I}^C$. We first perform the partial trace over subsystem $\ul{i}$, giving
    \begin{align}
        \sum_{\ul{i}, \ul{j} \in \ul{I}^C} \tr_{\ul{I}^C} \left( \ket{e_\ul{i}} \bra{e_\ul{j}} \right) &= \sum_{\ul{i}, \ul{j} \in \ul{I}^C} \tr_{\ul{I}^C \setminus \ul{i}} \left( \tr_{\ul{i}} \left( \ket{e_\ul{i}} \bra{e_\ul{j}} \right) \right) \nonumber \\
        &= \sum_{\ul{i}, \ul{j} \in \ul{I}^C} \tr_{\ul{I}^C \setminus \ul{i}} \left( \bra{0_{\ul{i}}} \left( \ket{e_\ul{i}} \bra{e_\ul{j}} \right) \ket{0_\ul{i}} + \bra{1_{\ul{i}}} \left( \ket{e_\ul{i}} \bra{e_\ul{j}} \right) \ket{1_\ul{i}}  \right) \, .
    \end{align}
    Then the first term vanishes as it is proportional to $\braket{0_{\ul{i}} \mid 1_{\ul{i}}}$, and the second term also vanishes unless $\ul{j} = \ul{i}$. Retaining this single non-zero contribution,
    \begin{align}
        \sum_{\ul{i}, \ul{j} \in \ul{I}^C} \tr_{\ul{I}^C} \left( \ket{e_\ul{i}} \bra{e_\ul{j}} \right) &= \sum_{\ul{i} = \ul{j} \in \ul{I}^C} \tr_{\ul{I}^C \setminus \ul{i}} \left( \bra{1_{\ul{i}}} \left( \ket{e_\ul{i}} \bra{e_\ul{j}} \right) \ket{1_\ul{i}}  \right) \nonumber \\
        &= k \ket{0^{\otimes m}} \bra{0^{\otimes m}} \, .
    \end{align}

    \item One of $\ul{i}$, $\ul{j}$ belongs to $\ul{I}$ while the other belongs to $\ul{I}^C$. Without loss of generality, assume that it is $\ul{i} \in \ul{I}^C$ and first perform the partial trace over $\ul{i}$. Then both terms
    \begin{align}
        \bra{0_{\ul{i}}} \left( \ket{e_\ul{i}} \bra{e_\ul{j}} \right) \ket{0_\ul{i}} + \bra{1_{\ul{i}}} \left( \ket{e_\ul{i}} \bra{e_\ul{j}} \right) \ket{1_\ul{i}}
    \end{align}
    vanish by orthogonality, the first because $\braket{0_{\ul{i}} \mid e_{\ul{i}}} = 0$ as before, and the second because $\braket{e_{\ul{j}} \mid 1_{\ul{i}}} = 0$ since by assumption $\ul{i} \neq \ul{j}$ so $e_{\ul{j}}$ has a tensor product factor of $\ket{0_{\ul{i}}}$ on subsystem $\ul{i}$. Therefore all such terms vanish:
    \begin{align}
        \sum_{\ul{i} \in \ul{I}, \ul{j} \in \ul{I}^C} \tr_{\ul{I}^C} \left( \ket{e_\ul{i}} \bra{e_\ul{j}} \right) = 0 = \sum_{\ul{i} \in \ul{I}^C, \ul{j} \in \ul{I}} \tr_{\ul{I}^C} \left( \ket{e_\ul{i}} \bra{e_\ul{j}} \right) \, .
    \end{align}

    \item Both $\ul{i}$ and $\ul{j}$ belong to $\ul{I}$. Then neither $\ket{e_\ul{i}}$ nor $\bra{e_\ul{j}}$ has a tensor product factor of $\ket{1}$ on any subsystem in the partial trace, and only the terms $\bra{0_{\ul{k}}} \left( \ket{e_\ul{i}} \bra{e_\ul{j}} \right) \ket{0_\ul{k}}$ contribute. Evaluating these partial traces has the effect of removing the extra $\ket{0}$ factors in $\ket{e_\ul{i}}$ and $\bra{e_\ul{j}}$, leaving the analogous states $\ket{f_{\ul{r}}}$ in $\mathcal{H}_{\ul{I}}$:
    \begin{align}
        \sum_{\ul{i}, \ul{j} \in \ul{I}} \tr_{\ul{I}^C} \left( \ket{e_\ul{i}} \bra{e_\ul{j}} \right) = \sum_{\ul{r}, \ul{s} \in \ul{I}} \ket{f_{\ul{r}}} \bra{f_{\ul{s}}} \, .
    \end{align}
\end{enumerate}
Combining the contributions from these three types of terms, we find
\begin{align}
    \rho^{\text{W}}_{\ul{I}} &= \tr_{\ul{I}^C} \left( \rho^{\text{W}}_{\ul{1} \ldots \ul{n}} \right) \nonumber \\
    &= \frac{k}{n} \ket{0^{\otimes m}} \bra{0^{\otimes m}} + \frac{1}{n} \sum_{\ul{r}, \ul{s} \in \ul{I}} \ket{f_{\ul{r}}} \bra{f_{\ul{s}}} \, ,
\end{align}
or in terms of the reduced W states
\begin{align}\label{nice_W_expr_reduced}
    \ket{ \text{W}_{ \ul{I} } } = \frac{1}{\sqrt{m}} \sum_{\ul{r} \in \ul{I}} \ket{f_\ul{r}} \, ,
\end{align}
one has
\begin{align}
    \rho^{\text{W}}_{\ul{I}} = \frac{k}{n} \ket{0^{\otimes m}} \bra{0^{\otimes m}} + \frac{m}{n} \ket{ \text{W}_{ \ul{I} } } \bra{ \text{W}_{ \ul{I} } } \, .
\end{align}
As a check, the trace of this reduced density matrix is unity, since $m + k = n$. 

It is then easy to read off the support projection for any such reduced density matrix:
\begin{align}
    s^{\text{W}}_{\ul{I}} = \ket{0^{\otimes m}} \bra{0^{\otimes m}} +  \ket{ \text{W}_{ \ul{I} } } \bra{ \text{W}_{ \ul{I} } } \, .
\end{align}
All that remains is to determine the dimensions of the kernels of the coboundary maps. A generic operator $\mathcal{O}_{\ul{I}}$ in $\mathrm{End} ( \mathrm{im} ( \rho^{\text{W}}_{\ul{I}} ) )$ has four non-trivial matrix elements:
\begin{align}\label{W_O_defn}
    \mathcal{O}_{\ul{I}} = a_{\ul{I}} \ket{ 0^{\otimes m} } \bra{ 0^{\otimes m}} + b_{\ul{I}} \ket{ 0^{\otimes m} } \bra{ \text{W}_{ \ul{I} } } + c_{\ul{I}} \ket{ \text{W}_{ \ul{I} } } \bra{ 0^{\otimes m}} + d_{\ul{I}} \ket{ \text{W}_{ \ul{I} } } \bra{ \text{W}_{ \ul{I} } } \, .
\end{align}
Like in the preceding proof, choose a subsystem $\ul{j} \notin \ul{I}$ and define the length $(m+1)$ multi-index $\ul{J} = \{ \ul{j} \} \cup \ul{I}$. We compute the matrix elements of terms relevant in the coboundary formula (\ref{d_defn}):
\begin{align}\label{W_d_basis_expansion}
    \bra{ 0^{\otimes ( m + 1 ) } } \iota_{\ul{I}}^{\ul{J}} ( \mathcal{O}_{\ul{I}} ) \ket{ 0^{\otimes ( m + 1 ) } } &= a_{\ul{I}} \, , \nonumber \\
    \bra{ 0^{\otimes ( m + 1 ) } } \iota_{\ul{I}}^{\ul{J}} ( \mathcal{O}_{\ul{I}} ) \ket{ \text{W}_{ \ul{J} } } &= \sqrt{ \frac{m}{m+1} } b_{\ul{I}} \, , \nonumber \\
    \bra{ \text{W}_{ \ul{J} } } \iota_{\ul{I}}^{\ul{J}} ( \mathcal{O}_{\ul{I}} ) \ket{ 0^{\otimes ( m + 1 ) } } &= \sqrt{ \frac{m}{m+1} } c_{\ul{I}} \, , \nonumber \\
    \bra{ \text{W}_{ \ul{J} } } \iota_{\ul{I}}^{\ul{J}} ( \mathcal{O}_{\ul{I}} ) \ket{ \text{W}_{ \ul{J} } } &= \frac{a_{\ul{I}}}{m+1} + \frac{m}{m+1} d_{\ul{I}} \, .
\end{align}
Unlike in the case of GHZ states, all four matrix elements are now non-trivial, and are in fact linearly independent. A given tuple of operators $\mathcal{O}_{\ul{I}}$ is therefore in the kernel of the coboundary map only if the alternating sum (\ref{alternating_constraint}) vanishes for each of the $a_{\ul{I}}$, $b_{\ul{I}}$, $c_{\ul{I}}$, and $d_{\ul{I}}$ separately. The dimension of each such solution space is ${n - 1 \choose m - 1}$, per Lemma \ref{simplex_lemma}, so
\begin{align}
    \dim ( \mathrm{ker} ( d^m ) ) = 4 {n - 1 \choose m - 1} \, ,
\end{align}
while using the rank-nullity theorem, $\dim \left( \Omega^m \right) = 4 {n \choose m}$, and Pascal's identity,
\begin{align}
    \dim ( \mathrm{im} ( d^m ) ) = 4 { n \choose m } - 4 { n - 1 \choose m - 1 } = 4 { n - 1 \choose m } \, .
\end{align}
The dimension of the cohomology is
\begin{align}
    \dim ( H^{m} ) &= \dim ( \ker ( d^{m} ) ) - \dim ( \mathrm{im} ( d^{m-1} ) ) \nonumber \\
    &= 4 \left( { n - 1 \choose m - 1 } - { n - 1 \choose m - 1 } \right) \nonumber \\
    &= 0 \, .
\end{align}
This is sufficient to prove that all of the interior cohomologies for the generalized W states vanish, again with the exception of the top and bottom cohomologies where the argument fails because the dimensions of those spaces of entanglement forms are $1$-dimensional rather than $2$-dimensional, which is why we restricted to $2 \leq k \leq n - 2$ above. We handle those remaining cases by hand. As $\dim ( \mathrm{im} ( d^0 ) ) = 1 = \dim ( \mathrm{im} ( d^{n-1} ) )$, we have
\begin{align}
    \dim ( H^1 ) &= \dim ( \ker ( d^1 ) ) - 1 \nonumber \\
    &= 4 { n \choose 0 } - 1 \nonumber \\
    &= 3 \, ,
\end{align}
and likewise
\begin{align}
    \dim ( H^{n-1} ) &= \dim ( \ker ( d^{n-1} ) ) - \dim ( \mathrm{im} ( d^{n-2} ) ) \nonumber \\
    &= \dim ( \Omega^{n-1} ) - 1- \left( \dim ( \Omega^{n-2} ) - \dim ( \ker ( d^{n-2} ) ) \right) \nonumber \\
    &= 4 n - 1 - 4 { n \choose n - 2 } + 4 { n - 1 \choose n - 3 } \nonumber \\
    &= 3 \, ,
\end{align}
where we have again used the rank-nullity theorem. This agrees with the Poincar\'e polynomial proposed in Conjecture \ref{W_conj}.
\end{proof}

\section{New Local Unitary Invariants}\label{sec:invariants}

When entanglement cohomology was first introduced in \cite{Mainiero:2019enr}, it was proven that the dimensions of these cohomologies are invariant under local unitary transformations acting on each of the tensor product factors $\mathcal{H}_{\ul{i}}$ of the total Hilbert space. Such LU transformations do not affect the structure of multipartite entanglement, so these Betti numbers are good quantities for characterizing inequivalent entanglement patterns, as we have seen in the case of the (unitarily inequivalent) generalized GHZ and W states.

However, two potential issues with the dimensions of entanglement cohomologies are:
\begin{enumerate}[label = (\Roman*)]
    \item\label{integer} The Betti numbers are \emph{integers}, which therefore cannot directly capture continuous data such as entanglement entropies or Schmidt coefficients. In a sense, they are measuring certain coarse-grained information about multipartite entanglement.

    \item\label{complete} The dimensions of cohomologies are not \emph{complete}, in the sense that there exist unitarily inequivalent pairs of pure states whose Betti numbers all coincide.
\end{enumerate}
The goal of this section is to introduce two new types of LU quantities, extending the collection of dimensions of cohomologies, which are constructed from the entanglement $k$-forms associated with a pure state. We take motivation from the close analogy between entanglement forms and differential forms on manifolds, defining quantities which are natural from a geometrical perspective. Both of the new classes of LU invariants are \emph{continuous} quantities, which resolves issue \ref{integer} of the dimensions of entanglement cohomologies.

It appears unlikely that even our extended collection of LU invariants is \emph{complete}, i.e. that this set of data is in one-to-one correspondence with unitarily inequivalent multipartite pure states, which would represent a solution to issue \ref{complete}. In general, performing a systematic classification of the possible invariants that can be constructed from objects transforming in representations of some group -- and the relations between such invariants -- is a difficult problem in the subject of invariant theory; see e.g. \cite{derksen2002computational} for an introduction. We will comment briefly on this future direction in Section \ref{sec:conclusion}. However, even if our set of new LU invariants is incomplete, these quantities still offer interesting new geometrically-inspired quantities which might be used to provide a partial classification of patterns of multipartite entanglement refining the one given by the Betti numbers.

\subsection{Spectrum of Entanglement Laplacian}\label{sec:lap}

We have emphasized that, given a pure state density matrix $\rho_{\ul{1} \ldots \ul{n}}$, the spaces of entanglement $k$-forms $\Omega^k ( \rho_{\ul{1} \ldots \ul{n}} )$ are simply length-${n \choose k}$ tuples of finite-dimensional linear operators. In particular, each such $\Omega^k ( \rho_{\ul{1} \ldots \ul{n}} )$ is itself a finite-dimensional vector space.
It was shown in \cite{Mainiero:2019enr} that any local unitary transformation induces an isomorphism of the vector spaces $\Omega^k ( \rho_{\ul{1} \ldots \ul{n}} )$, which can simply be viewed as a change of basis. The key ingredients of entanglement cohomology are constructed so that they transform appropriately with respect to this change of basis. For instance, owing to the definition of the coboundary operators $d^k$, any local unitary transformation simply induces a cochain isomorphism between the original entanglement cochain complex and the transformed one obtained by enacting the LU transformation. As a result, the dimensions of entanglement cohomologies are LU invariants, as we have reviewed.

Similarly, the entanglement Laplacian $\Delta$ is constructed only using the coboundary operators $d^k$ and their adjoints $\delta^k$ with respect to the LU-invariant inner product between entanglement forms. As a result, it is natural to conjecture that any collection of basis-independent data associated with the Laplacian -- such as the eigenvalues of the linear operator $\Delta$ acting on $\Omega^k ( \rho_{\ul{1} \ldots \ul{n}} )$, which are clearly basis-independent -- also provides LU invariants. We will not give an explicit mathematical proof of this statement, although we provide experimental checks of LU invariance of these eigenvalues below.

First let us make some simplistic comments. Like the Hodge Laplacian acting on differential forms, the entanglement Laplacian $\Delta$ is positive semi-definite, since
\begin{align}
    \langle \omega , \Delta \omega \rangle &= \langle \omega, d \delta \omega \rangle + \langle \omega, \delta d \omega \rangle \nonumber \\
    &= \langle \delta \omega , \delta \omega \rangle + \langle d \omega, d \omega \rangle \nonumber \\
    &= \left\| \delta \omega \right\|^2 + \left\| d \omega \right\|^2 \nonumber \\
    &\geq 0 \, ,
\end{align}
since the inner product on entanglement $k$-forms obeys the usual property that $\langle \eta, \eta \rangle \geq 0$ with equality only if $\eta = 0$. Therefore, the eigenvalues of $\Delta$ are non-negative: if $\Delta \omega = \lambda \omega$, then $\lambda \geq 0$, with $\lambda = 0$ only for harmonic forms.

Second, by virtue of its construction, the entanglement Laplacian commutes with differential and codifferential operators:
\begin{align}
    [ \Delta, d ] &= \left( d \delta + \delta d \right) d - d \left( d \delta + \delta d \right) \nonumber \\
    &= d \delta d - d \delta d \nonumber \\
    &= 0 \, ,
\end{align}
using the property that $d^2 = 0$, and likewise
\begin{align}
    [ \Delta, \delta ] &= \left( d \delta + \delta d \right) \delta - \delta \left( d \delta + \delta d \right) \nonumber \\
    &= \delta d \delta - \delta d \delta \nonumber \\
    &= 0 \, .
\end{align}
Thus if an entanglement $k$-form $\omega$ is an eigenform of $\Delta$ with eigenvalue $\lambda$, there is typically either a corresponding $(k+1)$-form $d \omega$ or a $(k-1)$-form $\delta \omega$ with the same eigenvalue, except in the special case of harmonic forms $\omega$ for which $d \omega = 0 = \delta \omega$. This is the same structure that underlies the pairing of energy levels in supersymmetric quantum mechanics \cite{Witten:1982csb,Witten:1982mt}, which applies to excited states but not to zero-energy ground states.

Therefore, if we consider the collection of \emph{all} non-zero eigenvalues of $\Delta$ as it acts on entanglement $k$-forms across all choices of $k$, there will be a two-fold degeneracy, as each eigenvalue is paired with the corresponding eigenvalue for its superpartner.

\subsubsection*{\ul{\it Numerical Experiments}}

The goal of the present section is to numerically investigate the spectrum of the entanglement Laplacian in simple examples, including states of two qubits and generalized GHZ and W states on three qubits.

The simplest examples to begin with are a product state on two qubits in $\mathcal{H}_{AB}$,
\begin{align}
    \ket{\psi^{(P)}} = \ket{0_A 0_B} \, ,
\end{align}
and an entangled state of two qubits,
\begin{align}
    \ket{\psi^{(E)}} = \frac{1}{\sqrt{2}} \left( \ket{0_A 0_B} + \ket{1_A 1_B} \right) \, ,
\end{align}
where ${}^{(P)}$ is short for ``product'' and ${}^{(E)}$ for ``entangled''. We construct the corresponding density matrices
\begin{align}
    \rho^{(P)}_{AB} = \ket{\psi^{(P)}} \bra{\psi^{(P)}} \, , \qquad \rho^{(E)}_{AB} = \ket{\psi^{(E)}} \bra{\psi^{(E)}} \, ,
\end{align}
and the spaces of entanglement forms
\begin{align}
    \Omega^k ( \rho^{(X)}_{AB} ) \, , \qquad X \in \{ P, E \} \, .
\end{align}
Let $\Delta_k^{(X)}$ be the corresponding entanglement Laplacians, again for $X \in \{ P, E \} $, where we have restored the index $k$ to indicate which forms the operator acts on. The spaces of $0$-forms and top forms are always one-dimensional for any pure state entanglement complex, so the Laplacian simply acts as multiplication by a positive constant. Here we find
\begin{align}\label{spec_EP_02}
    \mathrm{spec} \left( \Delta_0^{(X)} \right) = ( 2 ) = \mathrm{spec} \left( \Delta_2^{(X)} \right) \, ,
\end{align}
for both the product state and the entangled state, where $\mathrm{spec}$ denotes the spectrum of eigenvalues (with multiplicity). In a bipartite system, we only have entanglement $k$-forms for $k = 0, 1, 2$, so the remaining non-trivial case is to study the spectrum of the Laplacian when acting on $1$-forms. We find
\begin{align}\label{spec_EP_1}
    \mathrm{spec} \left( \Delta_1^{(P)} \right) = ( 2 , 2 ) \, , \qquad \mathrm{spec} \left( \Delta_1^{(E)} \right) = ( 0 , 0 , 0 , 0 , 0 , 0 , 2 , 2 ) \, .
\end{align}
Note that the number of eigenvalues is different in the two cases, since the space of entanglement $1$-forms is $2 \left( \mathrm{rank} ( \rho^{(X)}_A ) \right)^2$, which is equal to $2$ for the product state (as the reduced density matrix is rank $1$) but equal to $8$ for the entangled state (since the reduced density matrix has rank $2$). As expected, the Laplacian for the entangled state has six zero modes, corresponding to the harmonic representatives for the six non-trivial cohomology classes. The Laplacian for the product state has no zero modes, as this state has trivial cohomology; entanglement cohomology is trivial for any global product state \cite{Mainiero:2019enr}.

As a test of local unitary invariance, we have repeated the numerical computations of the spectra (\ref{spec_EP_02}) and (\ref{spec_EP_1}) after first acting on the initial states using random Haar-distributed local unitaries $U_A$ and $U_B$ on the two subsystems. Among $10,000$ trials with different local unitaries acting on the two subsystems, the numerically computed spectra for the Laplacians agreed with those quoted above to within $10$ decimal places in all cases.

One can also consider the Schmidt decomposition of a bipartite state,
\begin{align}\label{two_qubit_interpolating}
    \ket{\Lambda} = \sqrt{\Lambda} \ket{0_A 0_B} + \sqrt{1 - \Lambda} \ket{1_A 1_B} \, ,
\end{align}
which represents the most general $2$-qubit state up to the action of local unitaries.\footnote{We use the symbol $\Lambda$, rather than the more common choice $\lambda$, for one of the Schmidt coefficients to avoid confusion with eigenvalues $\lambda$ of operators such as the entanglement Laplacian.}

We performed a numerical computation of the spectrum of the entanglement Laplacians for states $\ket{\Lambda}$ at $10,000$ equally-spaced values of $\Lambda$, ranging from $\Lambda = 0$ to $\Lambda = 1$, but excluding the endpoints. In all cases, the numerically computed spectra agreed with that of the entangled state $\ket{\psi^{(E)}}$ to within ten decimal places. We therefore conclude that, at least for the simple case of $2$-qubit states, the spectrum of the Laplacian is not sensitive to the precise details of continuous quantities such as Schmidt coefficients.

More structure is apparent when we generalize to higher party number. For the $3$-party GHZ and W states, we find the spectra
\begin{align}\label{ghz3_spec}
    \mathrm{spec} \left( \Delta_0^{(\mathrm{GHZ}_3)} \right) &= ( 3 ) \, , \nonumber \\
    \mathrm{spec} \left( \Delta_1^{(\mathrm{GHZ}_3)} \right) &= ( 0, 0, 0, 0, 0, 0, 0, 3, 3, 3, 3, 3 ) \, , \nonumber \\
    \mathrm{spec} \left( \Delta_2^{(\mathrm{GHZ}_3)} \right) &= ( 0, 0, 0, 0, 0, 0, 0, 3, 3, 3, 3, 12 ) \, , \nonumber \\
    \mathrm{spec} \left( \Delta_3^{(\mathrm{GHZ}_3)} \right) &= ( 12 ) \, ,
\end{align}
and
\begin{align}\label{w3_spec}
    \mathrm{spec} \left( \Delta_0^{(\mathrm{W}_3)} \right) &= ( 3 ) \, , \nonumber \\
    \mathrm{spec} \left( \Delta_1^{(\mathrm{W}_3)} \right) &= ( 0, 0, 0, 0.75, 0.75, 1.5, 1.5, 1.5, 1.5, 3, 3, 3 ) \, , \nonumber \\
    \mathrm{spec} \left( \Delta_2^{(\mathrm{W}_3)} \right) &= ( 0, 0, 0, 0.75, 0.75, 1.5, 1.5, 1.5, 1.5, 3, 3, 12 ) \, , \nonumber \\
    \mathrm{spec} \left( \Delta_3^{(\mathrm{W}_3)} \right) &= ( 12 ) \, .
\end{align}
Again, both of these computations have been repeated $10,000$ times after first acting with a product of Haar-random local unitaries on the three subsystems and in all cases the result agrees with (\ref{ghz3_spec}) and (\ref{w3_spec}) up to ten decimal places, providing further numerical evidence for the LU invariance of the spectrum of the entanglement Laplacian.

A few comments are in order. First, the number of zero eigenvalues at each $k$ matches the dimension of the corresponding cohomology group $H^k$, as it must. Second, there is a pairing of energy levels at adjacent values of $k$ due to the fact that $\Delta$ commutes with $d$ and $\delta$; for instance, in both cases the top-forms with eigenvalue $12$ are paired with appropriate $2$-form ``superpartners'' that have the same eigenvalue. Third, unlike in the case of de Rham cohomology where under mild assumptions one finds that
\begin{align}
    [ \Delta_{\text{dR}}, \ast_{\text{dR}} ] = 0 \, ,
\end{align}
the Laplacian does \emph{not} commute with the Hodge star for entanglement cohomology, since otherwise the entanglement $2$-form $\omega$ with $\Delta \omega = 12 \omega$ would necessarily produce an associated entanglement $1$-form $\eta = \ast \omega$ obeying $\Delta \eta = 12 \eta$, but no such eigenform is observed.

\begin{figure}[htbp]
    \includegraphics[width=\linewidth]{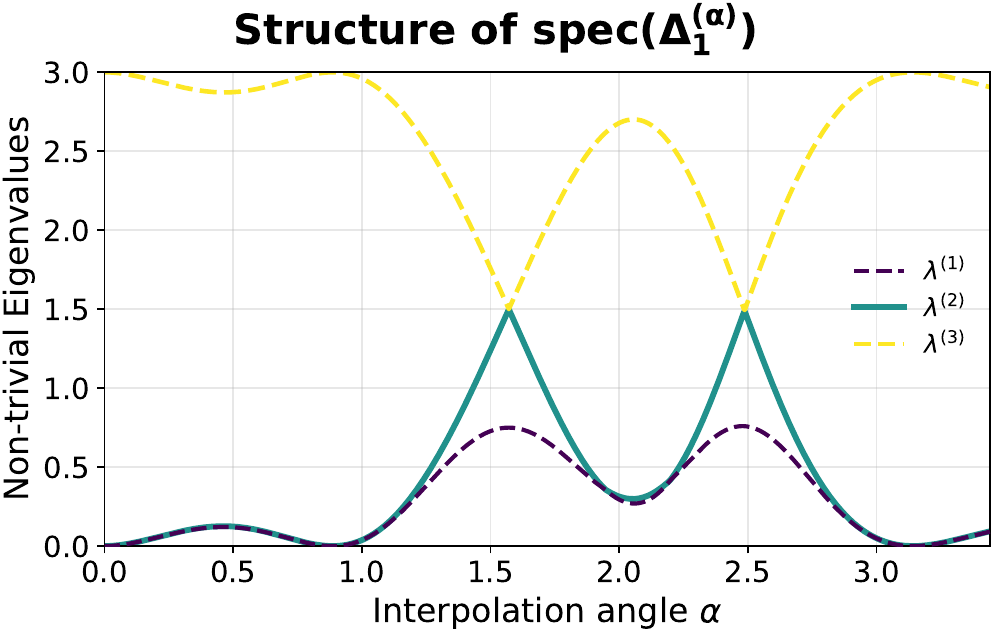}
    \caption{For each value of $\alpha$, we compute the eigenvalues of the entanglement Laplacian $\Delta_1^{(\alpha)}$ associated with the state (\ref{alpha_states}) that interpolates between $\ket{\mathrm{GHZ}_3}$ and $\ket{\mathrm{W}_3}$. Independently of $\alpha$, every such spectrum includes three zero eigenvalues and three eigenvalues equal to $3$, which represent the upper and lower bounds of the spectrum; we remove these and focus on the three intermediate eigenvalues $\lambda^{(1)} < \lambda^{(2)} < \lambda^{(3)}$. Each of these three non-trivial eigenvalues occurs with multiplicity two, but we plot only one such curve to represent each pair. Because we sort the eigenvalues in increasing order, the two level crossings in the plot appear as points where two eigenvalue curves meet in a cusp.}
    \label{fig:ghz_w_interp}
\end{figure}

Unlike in the bipartite case, where the entanglement Laplacian for the interpolating states (\ref{two_qubit_interpolating}) contained no additional data about the interpolation parameter $\Lambda$ (except at special points like $\Lambda = 0$ or $\Lambda = 1$ where the state becomes a product), there is a richer pattern in analogous tripartite states. Consider, for instance, the interpolating states
\begin{align}\label{alpha_states}
    \ket{\alpha} = \cos ( \alpha ) \ket{ \mathrm{GHZ}_3 } + \sin ( \alpha ) \ket{ \mathrm{W}_3 } \, .
\end{align}
We computed the spectrum of the entanglement Laplacians $\Delta_k^{(\alpha)}$ for various $\alpha \in ( 0, \pi )$. Since each of the eigenvalues in $\mathrm{spec} ( \Delta_2^{(\alpha)} )$ is paired with either one from $\mathrm{spec} ( \Delta_1^{(\alpha)} )$, or with the single eigenvalue of $\mathrm{spec} ( \Delta_2^{(\alpha)} ) = ( 12 )$, it suffices to focus on the spectrum of $\Delta_1^{(\alpha)}$ to see the structure. We find that this spectrum always includes the eigenvalues $0$ and $3$, each with multiplicity $3$, for all $\alpha$, so we remove these universal features to highlight the $\alpha$ dependence. The remaining six non-trivial eigenvalues come in three pairs, each with multiplicity two, which vary with $\alpha$; these three curves are displayed in Figure \ref{fig:ghz_w_interp}. 

In this setting, we see several phenomena -- including level crossing, local maxima and minima in eigenvalues, and points of tangency -- which were not present in the bipartite case. This demonstrates that there is strictly more information in the spectrum of the entanglement Laplacian for general states than is contained in the Poincar\'e polynomials.

Finally, let us comment that one could continue in this way, computing the spectra of Laplacians for higher $\ket{ \mathrm{GHZ}_n }$ and $\ket{ \mathrm{W}_n }$ states. We will adopt the compact notation $( \lambda : m )$ to indicate that an eigenvalue $\lambda$ is repeated with multiplicity $m$. 

For $n = 4$, we find
\begin{align}
    \mathrm{spec} \left( \Delta_0^{(\mathrm{GHZ}_4)} \right) &= ( 4 : 1 ) \, , \nonumber \\
    \mathrm{spec} \left( \Delta_1^{(\mathrm{GHZ}_4)} \right) &= ( 0 : 9, 4 : 7 ) \, , \nonumber \\
    \mathrm{spec} \left( \Delta_2^{(\mathrm{GHZ}_4)} \right) &= ( 0: 12, 4 : 6, 16 : 6 ) \, , \nonumber \\
    \mathrm{spec} \left( \Delta_3^{(\mathrm{GHZ}_4)} \right) &= ( 0 : 9, 16 : 6, 36 : 1 ) \, , \nonumber \\
    \mathrm{spec} \left( \Delta_4^{(\mathrm{GHZ}_4)} \right) &= ( 36 : 1 ) \, , 
\end{align}
and
\begin{align}
    \mathrm{spec} \left( \Delta_0^{(\mathrm{W}_4)} \right) &= ( 4 : 1 ) \, , \nonumber \\
    \mathrm{spec} \left( \Delta_1^{(\mathrm{W}_4)} \right) &= ( 0 : 3, 1.333: 3, 2.309: 6, 4 : 4 ) \, , \nonumber \\
    \mathrm{spec} \left( \Delta_2^{(\mathrm{W}_4)} \right) &= ( 1.333: 3, 2.309: 6, 4.0: 3, 5.333: 3, 9.238: 6, 16.0: 3 ) \, , \nonumber \\
    \mathrm{spec} \left( \Delta_3^{(\mathrm{W}_4)} \right) &= ( 0 : 3, 5.333: 3, 9.238: 6,   16 : 3, 36 : 1 ) \, , \nonumber \\
    \mathrm{spec} \left( \Delta_4^{(\mathrm{W}_4)} \right) &= ( 36 : 1 ) \, ,
\end{align}
and at $n = 5$ one observes
\begin{align}
    \mathrm{spec} \left( \Delta_0^{(\mathrm{GHZ}_5)} \right) &= ( 5 : 1 ) \, , \nonumber \\
    \mathrm{spec} \left( \Delta_1^{(\mathrm{GHZ}_5)} \right) &= ( 0 : 11, 5 : 9 ) \, , \nonumber \\
    \mathrm{spec} \left( \Delta_2^{(\mathrm{GHZ}_5)} \right) &= ( 0 : 20,  5 : 8, 20 : 12 ) \, , \nonumber \\
    \mathrm{spec} \left( \Delta_3^{(\mathrm{GHZ}_5)} \right) &= ( 0 : 20, 20 : 12, 45 : 8 ) \, , \nonumber \\
    \mathrm{spec} \left( \Delta_4^{(\mathrm{GHZ}_5)} \right) &= ( 0 : 11, 45 : 8, 80 : 1 ) \, , \nonumber \\
    \mathrm{spec} \left( \Delta_5^{(\mathrm{GHZ}_5)} \right) &= ( 80 : 1 ) \, , 
\end{align}
and
\begin{align}
    \mathrm{spec} \left( \Delta_0^{(\mathrm{W}_5)} \right) &= ( 5 : 1 ) \, , \nonumber \\
    \mathrm{spec} \left( \Delta_1^{(\mathrm{W}_5)} \right) &= ( 0 : 3, 1.875: 4, 3.062 : 8, 5 : 5 ) \, , \nonumber \\
    \mathrm{spec} \left( \Delta_2^{(\mathrm{W}_5)} \right) &= ( 1.875: 4, 3.062: 8,  5 : 4, 8.889: 6, 13.333: 12,   20 : 6 ) \, , \nonumber \\
    \mathrm{spec} \left( \Delta_3^{(\mathrm{W}_5)} \right) &= ( 8.889: 6, 13.333: 12, 16.875: 4, 20 : 6, 27.557: 8, 45 : 4 ) \, , \nonumber \\
    \mathrm{spec} \left( \Delta_4^{(\mathrm{W}_5)} \right) &= ( 0 : 3, 16.875: 4, 27.557: 8,  45 : 4,  80 : 1 ) \, , \nonumber \\
    \mathrm{spec} \left( \Delta_5^{(\mathrm{W}_5)} \right) &= ( 80 : 1 ) \, .
\end{align}
It appears that $\Delta_0$ always acts as multiplication by $n$ while $\Delta_n$ always acts as multiplication by $n ( n - 1 )^2$, although the structure of the other eigenvalues seems more intricate. It would be very interesting if one could conjecture and prove exact closed-form expressions for the spectra of these operators across all values of $n$, analogous to those for the Poincar\'e polynomials which we established in Section \ref{sec:proofs}.

\subsubsection*{\ul{\it K\"unneth Theorem}}

As a final application of the entanglement Laplacian, let us comment that it allows us to easily give a new proof of a (known) K\"unneth theorem for entanglement cohomology of pure states, using different techniques than the discussion in Section 7.6 of \cite{Mainiero:2019enr}. In general, K\"unneth theorems relate the homology (or cohomology) of a product of objects to the homologies of the objects themselves. The most familiar case might be the formula for singular homology with coefficients in a field $F$, which relates the homology groups $H_i$ of topological spaces $X$, $Y$, and their product $X \times Y$ as
\begin{align}\label{default_kunneth}
    H_k ( X \times Y ; F ) \cong \bigoplus_{i + j = k} H_i ( X; F ) \otimes H_j ( Y; F ) \, .
\end{align}
Taking dimensions on both sides of the formula (\ref{default_kunneth}) then gives relations for the dimensions of homologies, i.e. for their Poincar\'e polynomials.

We will focus on the dimensions of entanglement cohomologies associated with tensor products of pure states, giving the argument only in brief schematic form since it closely parallels the corresponding proof for manifolds. Consider two pure states $\ket{\psi}$ and $\ket{\chi}$ on disjoint collections of subsystems $\ul{I} = \{ \ul{1}, \ldots, \ul{n} \}$ and $\ul{J} = \{ \ul{n+1}, \ldots, \ul{n + m} \}$, along with their associated pure state density matrices $\rho_{\ul{I}}$ and $\rho_{\ul{J}}$. Let $\Delta_{\psi \otimes \chi}$ be the entanglement Laplacian acting on $(k+p)$-forms associated with the pure state $\ket{\psi} \otimes \ket{\chi}$ on the full set of subsystems $\ul{1}, \ldots, \ul{n + m}$, and let $\Delta_\psi$, $\Delta_\chi$ be the individual Laplacians acting on entanglement forms $\Omega^k ( \rho_{\ul{I}} )$ and $\Omega^p ( \rho_{\ul{J}} )$, respectively. Given two such entanglement forms
\begin{align}
    \alpha \in \Omega^k ( \rho_{\ul{I}} ) \, , \qquad \beta \in \Omega^p ( \rho_{\ul{J}} ) \, ,
\end{align}
one can also naturally identify them as entanglement forms associated with the full tensor product system, simply setting components to zero if they involve any indices belonging to the complement system. Making this identification, one finds that
\begin{align}
    \Delta_{\psi \otimes \chi} \left( \alpha \wedge \beta \right) = 0 \, ,
\end{align}
if and only if
\begin{align}
    \Delta_\psi \alpha = 0 = \Delta_\chi \beta \, .
\end{align}
This follows from the facts about the Hodge star, wedge product, and inner product introduced in \cite{Ferko:2024swt}, such as the fact that the exterior derivative satisfies a Leibniz rule with respect to the wedge product. 

We write this schematically as
\begin{align}
    \mathrm{Harm}^q ( \psi \otimes \chi ) = \bigoplus_{k + p = q } \mathrm{Harm}^k ( \psi ) \otimes \mathrm{Harm}^p ( \chi ) \, ,
\end{align}
where $\mathrm{Harm}^q$ denotes the space of harmonic entanglement $q$-forms, which is isomorphic to the $q$-th cohomology group by the Hodge isomorphism theorem of \cite{Ferko:2024swt}, and thus
\begin{align}
    H^q ( \psi \otimes \chi ) = \bigoplus_{i + j = q} H^i ( \psi ) \otimes H^j ( \chi ) \, .
\end{align}
Taking the dimensions of these cohomology groups and defining the corresponding Betti numbers $b^k_{\psi \otimes \chi}$, $b^i_\psi$, and $b^j_\chi$, we also see that
\begin{align}\label{betti_product_result}
    b^q_{\psi \otimes \chi} = \sum_{i + j = q} b^i_\psi b^j_\chi \, .
\end{align}
Combining the result (\ref{betti_product_result}), proven either using the general homological arguments of \cite{Mainiero:2019enr} or the use of the entanglement Laplacian, with the results of Section \ref{sec:proofs}, then gives a rigorous derivation of the Betti numbers for pure states that can be written as arbitrary tensor products of generalized GHZ and W states. For instance, states of the form
\begin{align}\label{GHZ_W_product}
    \ket{ \mathrm{GHZ}_{n, \vec{d} } } \otimes \ket{ \text{W}_{n', \vec{d}' } }
\end{align}
have Poincar\'e polynomials determined by (\ref{betti_product_result}) and the (now proven) expressions for the Betti numbers of GHZ and W states given in Conjectures \ref{GHZ_conj} and \ref{W_conj}. Tensor products of three or more such states have Poincar\'e polynomials given by iteratively applying (\ref{betti_product_result}).

\begin{table}[h]
  \centering
\begin{tabular}{ |p{1.1cm}||p{4.1cm}|p{4.1cm}|p{4.1cm}|  }
 \hline
 \multicolumn{4}{|c|}{Numerical Poincar\'e Polynomials of States $\ket{ \mathrm{GHZ}_{n }} \otimes \ket{ \text{W}_{m }}$} \\
 \hline
 \diagbox{$n$}{$m$} & \multicolumn{1}{c|}{$3$}
     & \multicolumn{1}{c|}{$4$} & \multicolumn{1}{c|}{$5$} \\
 \hline
  $3$ & $21 y + 42 y^2 + 21 y^3$ & $21 y + 21 y^2 + 21 y^3 + 21 y^4$ & $21 y + 21 y^2 + 21 y^4 + 21 y^5$ \\ \hline
  $4$ & $27 y + 63 y^2 + 63 y^3 + 27 y^4$  & $27 y + 36 y^2 + 54 y^3 + 36 y^4 + 27 y^5$ & $27 y + 36 y^2 + 27 y^3 + 27 y^4 + 36 y^5 + 27 y^6$ \\ \hline
 $5$ & $33 y + 93 y^2 + 120 y^3 + 93 y^4 + 33 y^5$ & $33 y + 60 y^2 + 93 y^3 + 93 y^4 + 60 y^5 + 33 y^6$ & $33 y + 60 y^2 + 60 y^3 + 66 y^4 + 60 y^5 + 60 y^6 + 33 y^7$ \\
 \hline
\end{tabular}
  \caption{Experimental values of the Poincar\'e polynomials for tensor products of generalized GHZ and W states are computed for various numbers of subsystems. The dimensions of cohomologies are evaluated using the same methodology described in the caption of Table \ref{tab:ghz-w-poincare}, i.e. representing the designated tensor product state as a QuTiP \cite{Johansson:2011jer,Johansson:2012qtx} object and computing the images and kernels of coboundary operators $d^k$ using numerical linear algebra. All results are consistent with the prediction from the K\"unneth formula and the Poincar\'e polynomials for generalized GHZ and W states given in Conjectures \ref{GHZ_conj} and \ref{W_conj}, which were proven in Sections \ref{sec:ghz_proof} and \ref{sec:w_proof}, respectively.}
  \label{tab:tensor-product-poincare}
\end{table}

We have numerically computed the Poincar\'e polynomials for several tensor products of the form (\ref{GHZ_W_product}), which are displayed in Table \ref{tab:tensor-product-poincare} and all take the expected form.

\subsection{Intersection Numbers}

We now turn to the second class of new invariants of interest in this work, which are inspired by certain integrals that measure intersections of submanifolds in geometry.

\subsubsection*{\ul{\it Review and Motivation}}

In general, the advantage of having a cohomology \emph{ring} as opposed to mere cohomology \emph{groups} is that the ring structure offers additional information that can sometimes be used to distinguish between objects which have identical cohomology groups. For instance, in singular cohomology, the ring structure is provided by the \emph{cup product}: given a topological space $X$ and commutative ring $R$, one can define a map
\begin{align}
    H^p ( X, R ) \times H^q ( X, R ) \to H^{p + q} ( X, R ) 
\end{align}
that bilinearly maps a $p$-cochain and a $q$-cochain to a $(p+q)$-cochain. This product endows the collection of cohomology groups with the structure of a graded ring. In ordinary de Rham cohomology, the role of the cup product is played by the wedge product of differential forms, and in entanglement cohomology, the appropriate version of the cup product is defined in equation (\ref{wedge_definition}).

In the setting of manifolds, another useful feature of the cohomology ring is that it allows us to study \emph{intersections} algebraically, using Poincar\'e duality. On an oriented closed manifold $M$ of dimension $n$, for every closed $p$-form $\omega$ there is an $(n-p)$-cycle $N$ such that
\begin{align}
    \int_{M} \omega \wedge \eta = \int_{N} \eta \, ,
\end{align}
for all closed $(n-p)$-forms $\eta$. This yields an isomorphism between the cohomology groups $H^p ( M )$ and the corresponding homology groups $H_{n - p} ( M )$. Under this isomorphism, the wedge product (or cup product) is dual to the intersection of submanifolds. Consider two closed, oriented, embedded submanifolds $A$ and $B$ of the parent manifold $M$, with dimensions $p$ and $q$, respectively, such that $p + q = n$. Under the technical assumption that $A$ and $B$ intersect transversally, which means that their tangent spaces obey $T_x A + T_x B = T_x M$ at each point $x \in A \cap B$, the submanifolds $A$ and $B$ intersect at a finite collection of points $x$, and one can count each such point with a sign $\epsilon_x$ that measures the orientation of the tangent spaces $T_x A$ and $T_x B$ at $x$. One then defines the intersection number,
\begin{align}
    A \cdot B = \sum_{x \in A \cap B} \epsilon_x \, .
\end{align}
As we mentioned, Poincar\'e duality gives a relationship between closed $p$-forms and $(n-p)$-cycles. If we wish to compute the intersection number of two cycles $A$ and $B$ -- perhaps first deforming them to other cycles within the same homology classes, to ensure that they intersect transversally -- we can first find a closed $q$-form $\omega_A$ which is Poincar\'e dual to $A$ (as $q = n - p$), and a closed $p$-form $\omega_B$ representing the Poincar\'e dual of $B$. The interesting fact is that the intersection number is related to these forms as
\begin{align}
    A \cdot B = \int_M \omega_A \wedge \omega_B \, .
\end{align}
This allows us to relate a geometrical quantity, the intersection number, to an algebraic integral of the wedge product of differential forms, which is easier to compute.

Such intersection numbers appear in several contexts in physics. For instance, in string compactifications, one is often interested in Calabi-Yau $3$-folds, which are manifolds $M$ of real dimension six. Given a collection of $4$-cycles (or divisor classes) $D_i$, one can likewise identify their Poincar\'e dual $2$-forms $\omega_i$, and compute the triple intersection numbers
\begin{align}
    \kappa_{ijk} = \int_M \omega_i \wedge \omega_j \wedge \omega_k \, ,
\end{align}
which measures a signed count of the finite collection of intersection points $D_i \cap D_j \cap D_k$. In type IIA compactifications on a Calabi–Yau threefold, $\kappa_{ijk}$ give the classical cubic terms in the holomorphic prepotential for the complexified Kähler moduli, and the corresponding Yukawa couplings are encoded by its third derivatives (up to instanton corrections).

\subsubsection*{\ul{\it Definition and Basic Properties}}

Motivated by the utility of this framework, we wish to define and study similar quantities in the setting of entanglement cohomology. We have already defined natural analogues of differential forms on manifolds and their wedge product, and the role of integration over the parent manifold $M$ is played by a trace over the full multi-partite Hilbert space. Therefore, given a collection of closed entanglement $k_a$-forms
\begin{align}\label{alpha_ki_defn}
    \alpha^{(k_1)} \in \Omega^{k_1} \left( \rho_{\ul{1} \ldots \ul{n}} \right) \, , \ldots \, , \alpha^{(k_m)}  \in \Omega^{k_m} \left( \rho_{\ul{1} \ldots \ul{n}} \right) \, ,
\end{align}
such that $\sum_a k_a = n$, we will define their intersection pairing by
\begin{align}\label{intersection_pairing}
    I \left( \alpha^{(k_1)} \, , \ldots \, , \alpha^{(k_m)} \right) = \tr_{\ul{1} \ldots \ul{n}} \left( \alpha^{(k_1)} \wedge \ldots \wedge  \alpha^{(k_m)} \right) \, .
\end{align}
In order to eliminate the ambiguity to shift the closed forms $\alpha^{(k_1)}$ by exact forms, we will typically choose them to be representatives of cohomology classes, and for concreteness it is often convenient to use the unique harmonic representative of each such class.

More specifically, we now take an indexed family of the collection of closed forms (\ref{alpha_ki_defn}), where each one is harmonic and belongs to a distinct cohomology class:
\begin{align}
    \alpha^{(k_a)}_{i_a} \in H^{k_a} \left( \rho_{\ul{1} \ldots \ul{n}} \right) \, , \qquad \Delta \alpha^{(k_a)}_{i_a} = 0 \, ,
\end{align}
where now $i_a = 1 , \ldots , \dim \left( H^{k_a} \left( \rho_{\ul{1} \ldots \ul{n}} \right) \right)$ for each $a = 1 , \ldots, m$. Demanding that each $\alpha^{(k_a)}_{i_a}$ belong to a different class within $H^{k_a}$ eliminates all ambiguity in the choice of differential form, and the value of the trace (\ref{intersection_pairing}) is automatically unambiguous because each such form is closed. We now form the indexed collection of quantities
\begin{align}\label{intersection_numbers}
    M_{i_1 \ldots i_m} = I \left( \alpha^{(k_1)}_{i_1} , \ldots , \alpha^{(k_m)}_{i_m} \right) \, ,
\end{align}
which we call the intersection numbers. We will sometimes adopt more specific terminology like intersection matrix when $m = 2$, or triple intersections when $m = 3$.

Although the quantities (\ref{intersection_numbers}) are unambiguously defined for any fixed choice of basis for the cohomology classes, as $M_{i_1 \ldots i_m}$ carries indices, it will transform covariantly under changes of basis for each of the vector spaces $H^{k_a}$. To eliminate this ambiguity and define fully invariant quantities, we should form singlets by taking products of the $M_{i_1 \ldots i_m}$ and contracting all of the indices. For instance, one could compute the quadratic invariant
\begin{align}\label{example_invariant}
    M_{i_1 \ldots i_m} M^{i_1 \ldots i_m} \, ,
\end{align}
which is a scalar under rotations of the bases for any of the $H^{k_a}$. The collection of such scalar quantities like (\ref{example_invariant}), which are constructed from intersection numbers but are singlets under rotations, are the new invariants that we propose to study.

For instance, in the case of the intersection matrix
\begin{align}
    M_{ij} = I \left( \alpha_i^{(k_1)} , \alpha_j^{(k_2)} \right) \, , \qquad k_1 + k_2 = n \, , 
\end{align}
an obvious collection of basis-independent scalars to study are its eigenvalues.

For general tensors $M_{i_1 \ldots i_m}$ with many indices, it is not known how to systematically enumerate all independent invariants like (\ref{example_invariant}) that can be constructed from $M_{i_1 \ldots i_m}$. This is the question addressed in the mathematical subject of invariant theory, which we have already mentioned. We will not attempt to address this difficult problem here, preferring instead to select a few simple invariants to compute in the examples below.

As in the case of the spectrum of the entanglement Laplacian, since the intersection numbers are constructed only from ingredients within entanglement cohomology that transform naturally under local unitary transformations, it is reasonable to expect that all scalars built from (\ref{intersection_numbers}) are LU invariants. Indeed, an argument for this conclusion is the following. Consider a transformation by a tensor product of local unitary transformations,
\begin{align}\label{local_unitary}
    U_{\ul{1} \ldots \ul{n}} = U_{\ul{1}} \otimes \ldots \otimes U_{\ul{n}} \, .
\end{align}
A local operator $\mathcal{O}_{\ul{i}}$ acting on a Hilbert space $\mathcal{H}_{\ul{i}}$ transforms as
\begin{align}\label{U_one_subsystem}
    \mathcal{O}_{\ul{i}} \longrightarrow U_{\ul{i}} \mathcal{O}_{\ul{i}} U^\dagger_{\ul{i}} \, .
\end{align}
Now let $\alpha \in \Omega^k ( \rho_{\ul{1} \ldots \ul{n}} )$ be an entanglement $k$-form, which can be viewed as a tuple of operators acting on $k$-fold tensor products of the local Hilbert spaces (suitably restricted to images of reduced density matrices). Let $U_{\ul{I}} \cdot \alpha_{\ul{I}}$ represent the action of the local unitary (\ref{local_unitary}) on a component of $\alpha$, where $\ul{I}$ is a length-$k$ multi-index, which simply acts as (\ref{U_one_subsystem}) on each subsystem $\ul{i} \in \ul{I}$, and let $U \cdot \alpha$ represent the full tuple of such transformed components as $\ul{I}$ ranges over all length-$k$ multi-indices. Because the transformation (\ref{local_unitary}), by assumption, acts independently on each tensor product factor, by the definition of the wedge product (\ref{wedge_definition}) one has
\begin{align}
    U \cdot \left( \omega \wedge \eta \right) = \left( U \cdot \omega \right) \wedge \left( U \cdot \eta \right) \, .
\end{align}
Therefore, under the action of a local unitary, the intersection pairing (\ref{intersection_pairing}) transforms as
\begin{align}\label{intersection_transformed}
    I \left( \alpha^{(k_1)} \, , \ldots \, , \alpha^{(k_m)} \right) &\longrightarrow I \left( U \cdot \alpha^{(k_1)} \, , \ldots \, , U \cdot \alpha^{(k_m)} \right) \nonumber \\
    &= \tr_{\ul{1} \ldots \ul{n}} \left( \left( U \cdot \alpha^{(k_1)} \right) \wedge \ldots \wedge  \left( U \cdot \alpha^{(k_m)} \right) \right) \nonumber \\
    &= \tr_{\ul{1} \ldots \ul{n}} \left( U \cdot \left( \alpha^{(k_1)} \wedge \ldots \wedge \alpha^{(k_m)} \right) \right) \nonumber \\
    &= \tr_{\ul{1} \ldots \ul{n}} \left( U_{\ul{1} \ldots \ul{n}} \left( \alpha^{(k_1)} \wedge \ldots \wedge \alpha^{(k_m)} \right) U_{\ul{1} \ldots \ul{n}}^\dagger \right) \nonumber \\
    &= \tr_{\ul{1} \ldots \ul{n}} \left( \alpha^{(k_1)} \wedge \ldots \wedge \alpha^{(k_m)} \right) \nonumber \\
    &= I \left( \alpha^{(k_1)} \, , \ldots \, , \alpha^{(k_m)} \right) \, ,
\end{align}
where we have used that the unitary transformation acts as $\mathcal{O}_{\ul{1} \ldots \ul{n}} \to U_{\ul{1} \ldots \ul{n}} \mathcal{O}_{\ul{1} \ldots \ul{n}} U_{\ul{1} \ldots \ul{n}}^\dagger$ for an operator on the total Hilbert space $\mathcal{H}_{\ul{1} \ldots \ul{n}}$, and cyclicity of the trace.

Below we will give further evidence for LU invariance by relating some scalars built from intersection numbers to other LU-invariant quantities such as Schmidt coefficients.

\subsubsection*{\ul{\it Numerical Experiments}}

As an example, we consider the general Schmidt form $\ket{\Lambda}$ of a two-qubit state introduced in equation (\ref{two_qubit_interpolating}), for $0 < \Lambda < 1$. All of these states have a first entanglement cohomology $H^1$ of dimension $6$, and again for concreteness, let $\alpha_i^{(\Lambda)}$ be the unique harmonic representatives of these cohomology classes. We define the intersection matrix
\begin{align}\label{intersection_matrix}
    M_{ij}^{(\Lambda)} = \tr_{AB} \left( \alpha_i^{(\Lambda)} \wedge \alpha_j^{(\Lambda)} \right) \, .
\end{align}
As we have commented, the entries of $M_{ij}^{(\Lambda)}$ are not invariant under rotations of the basis for the harmonic $1$-forms. However, the eigenvalues of this intersection matrix are basis-independent, so we consider the spectrum $\mathrm{spec} \left( M_{ij}^{(\Lambda)} \right)$. Since the wedge product of odd $k$-forms is antisymmetric, $M_{ij}^{(\Lambda)}$ is an antisymmetric $6 \times 6$ matrix -- which we find numerically to be real -- and therefore has purely imaginary eigenvalues which come in complex-conjugate pairs. Experimentally, we find that every such spectrum takes the form
\begin{align}\label{spec_intersection}
    \mathrm{spec} \left( M_{ij}^{(\Lambda)} \right) = \left( i x^{(\Lambda)} , - i x^{(\Lambda)} , i y^{(\Lambda)} , - i y^{(\Lambda)} , i y^{(\Lambda)} , - i y^{(\Lambda)} \right) \, ,
\end{align}
for two positive quantities $x^{(\Lambda)}$ and $y^{(\Lambda)}$. That is, among the three complex-conjugate pairs of eigenvalues, we find that one of the pairs is always distinct from the others, but the remaining two pairs are degenerate. 

\begin{figure}[htbp]
    \includegraphics[width=\linewidth]{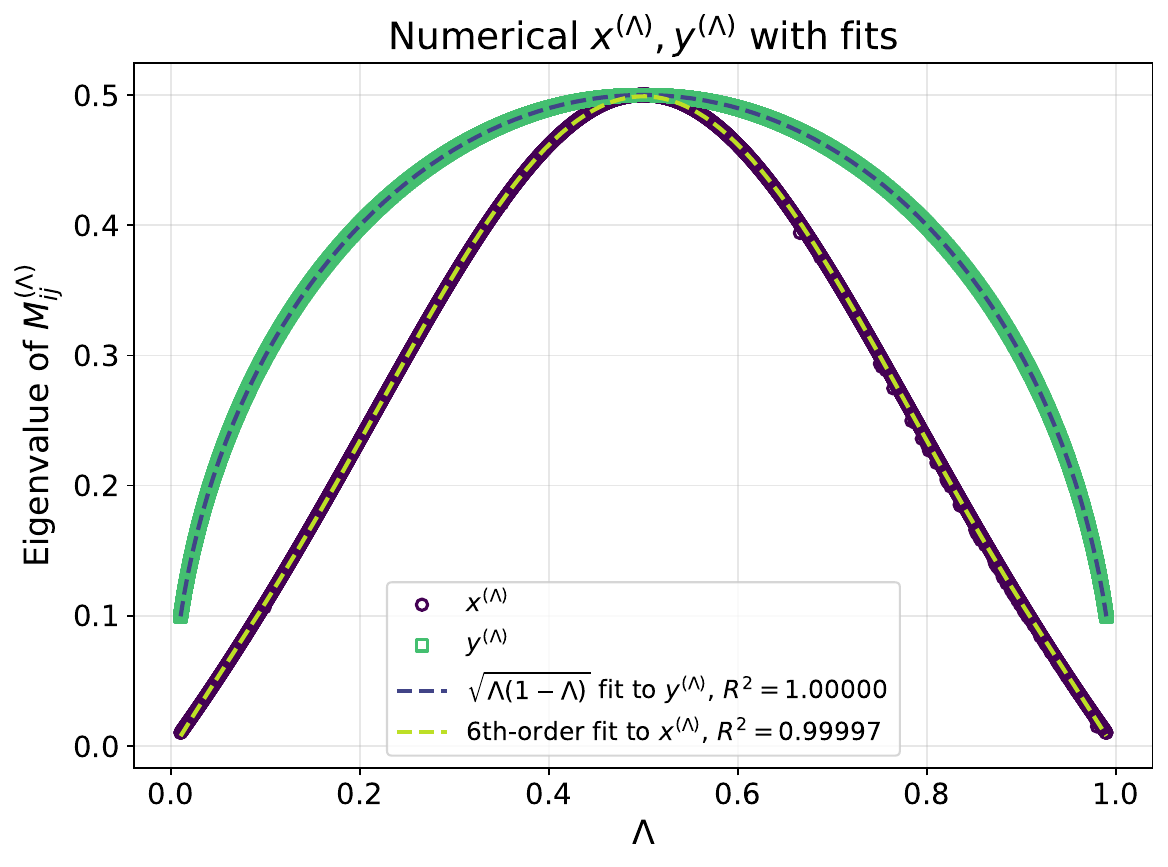}
    \caption{We display the experimental values of the eigenvalues $x^{(\Lambda)}$ and $y^{(\Lambda)}$ of the matrix $M_{ij}^{(\Lambda)}$ given in equation (\ref{spec_intersection}). For each value of the Schmidt coefficient $\Lambda$, we prepare the state (\ref{two_qubit_interpolating}), find a matrix representation for its entanglement Laplacian acting on entanglement $1$-forms, and then diagonalize to find the six independent $1$-forms which are in its kernel. These are harmonic forms $\alpha_i^{(\Lambda)}$ which give unique representatives for the six cohomology classes. We then compute the intersection matrix (\ref{intersection_matrix}) and numerically evaluate its eigenvalues, identifying $x^{(\Lambda)}$ and $y^{(\Lambda)}$ according to their multiplicities. These eigenvalues are plotted along with our proposed closed-form expression (\ref{yLambda_exact}) for $y^{(\Lambda)}$, while we perform a fit of $x^{(\Lambda)}$ to a sixth-order polynomial in $\Lambda$ and show the result.}
    \label{fig:lambda_figure}
\end{figure}

The numerically computed values of $x^{(\Lambda)}$ and $y^{(\Lambda)}$ for $10,000$ equally-spaced Schmidt coefficients $0 < \Lambda < 1$ are displayed in Figure \ref{fig:lambda_figure}. We find that the most degenerate eigenvalue agrees, with $R^2 = 1.00000$, with the exact expression
\begin{align}\label{yLambda_exact}
    y^{(\Lambda)} = \sqrt{ \Lambda ( 1 - \Lambda ) } \, ,
\end{align}
which is half the concurrence. We were unable to obtain a closed-form expression for the remaining eigenvalue $x^{(\Lambda)}$, but we display the result of a fit of a sixth-order polynomial in $\Lambda$ in Figure \ref{fig:lambda_figure}, with $R^2 = 0.99997$. The best fit polynomial takes the form
\begin{align}
    x^{(\Lambda)} = - 16.54 \Lambda^6 + 49.64 \Lambda^5 - 50.68 \Lambda^4 + 18.59 \Lambda^3 - 2.25 \Lambda^2 + 1.23 \Lambda \, ,
\end{align}
up to two decimal places.

As a consequence, if one has access to the intersection numbers for a bipartite system, it is possible to reconstruct the Schmidt coefficients by computing the eigenvalues of the intersection matrix, identifying the quantity $y^{(\Lambda)}$ associated with the degenerate eigenvalue, and inverting the relation (\ref{yLambda_exact}). This means that, unlike the dimensions of entanglement cohomologies and the spectrum of the entanglement Laplacian, the intersection matrix encodes data about Schmidt coefficients; in the case of a bipartite system, every LU invariant is a function of the Schmidt coefficients, and therefore the intersection matrix captures all LU information in this simple case. In particular, this provides further evidence that the intersection numbers are LU invariants, since at least in the bipartite case they are functions only of the Schmidt coefficients, which are themselves LU invariant.

Next let us consider more general multipartite systems. First, given an arbitrary pure state density matrix $\rho_{\ul{1} \ldots \ul{n}}$, we note that it is always possible to coarse-grain and interpret this as a bipartite system $\rho_{\ul{I} \ul{I}^C}$ on a collection of subsystems $\ul{I}$ and its complement. By repeating the analysis above, we find that computing the intersection matrix associated with this coarse-grained bipartite system gives access to the Schmidt coefficients. Therefore, the intersection numbers constructed using entanglement $k$-forms -- when combined with the coarse-graining operation -- also describes all Schmidt coefficients across bipartitions in an arbitrary multi-partite system.

However, one can extract more information from more general intersection pairings. For instance, in a tripartite pure state $\rho_{\ul{1} \ul{2} \ul{3}}$ let $\alpha_i^{(1)}$ be a basis for the first entanglement cohomology $H^1 ( \rho_{\ul{1} \ul{2} \ul{3} } )$, where $i = 1 , \ldots , \dim ( H^1 ( \rho_{\ul{1} \ul{2} \ul{3} } ) )$. For concreteness, we can again choose the $\alpha_i^{(1)}$ to be the unique harmonic representative of each cohomology class. One can then form the triple intersections
\begin{align}\label{triple_intersections}
    M_{ijk} = \tr_{\ul{1} \ul{2} \ul{3} } \left( \alpha_i^{(1)} \wedge \alpha_j^{(1)} \wedge \alpha_k^{(1)} \right) \, .
\end{align}
As before, one can then compute scalars which are invariant under rotations of the basis of the first cohomology, such as the contraction $M_{ijk} M^{ijk}$.\footnote{See Appendix B of \cite{Chandra:2023afu} for other invariants that can be constructed from triple intersections in the context of Calabi-Yau manifolds. See also \cite{Cederwall:2025ywy} for interesting recent work on the general theory of invariants.} Of course, in a more general $n$-partite system, one could also coarse-grain by partitioning the collection of subsystems into three disjoint subsets and then computing the triple intersection numbers (\ref{triple_intersections}). As the intersection matrix (\ref{intersection_matrix}) describes information about Schmidt coefficients across bipartitions, it is natural to expect that invariants constructed from the triple intersections should describe some analogous quantities which relate to tripartitions. It is well-known that there is no genuine multipartite generalization of the Schmidt decomposition in general -- see \cite{Peres:1994qv} for necessary and sufficient conditions that such a decomposition should exist -- but invariants constructed from triple (and higher) intersections might be interpreted, morally, as encoding some information of this form.

To conclude this subsection, let us investigate the properties of the invariant $M_{ijk} M^{ijk}$ constructed from the triple intersection numbers (\ref{triple_intersections}) for the $\ket{\alpha}$ states of (\ref{alpha_states}) which interpolate between $\ket{ \mathrm{GHZ}_3 }$ and $\ket{ \mathrm{W}_3 }$. Numerically, we find that
\begin{align}
    \left( M_{ijk} M^{ijk} \right)^{(0)} = 3
\end{align}
to within ten decimal places, which suggests that this scalar is equal to $3$ for the $\ket{\mathrm{GHZ}_3}$ state. However, for any other $\alpha \neq 0$, we find that 
\begin{align}
    \left( M_{ijk} M^{ijk} \right)^{(\alpha)} = 0 \, ,
\end{align}
again to at least ten decimal places. It therefore seems that this particular intersection number, somewhat like the three-tangle $\tau_3$, is some measure of genuine tripartite entanglement. As we have mentioned, $\ket{ \mathrm{GHZ}_3 }$ has ``purely tripartite entanglement'' in the sense that tracing out a single qubit leaves a classical mixture of product states with no bipartite entanglement, whereas the entanglement structure of the state $\ket{\mathrm{W}_3}$ includes bipartite entanglement, as tracing out one qubit from the W state leaves a system in which the two remaining qubits are still entangled. What is perhaps less obvious is that the invariant $M_{ijk} M^{ijk}$ does not smoothly depend on $\alpha$, but rather immediately drops to zero once the superposition involves any $\ket{\mathrm{W}_3}$ component, just as the dimensions of the entanglement cohomologies change discontinuously in the family of $\ket{\alpha}$ states.

It would be very interesting to investigate the structure of scalars constructed from higher intersection numbers in other families of multipartite states.

\section{Conclusion}\label{sec:conclusion}

In this work, we have explored and extended the formalism of entanglement cohomology in two primary directions. First, we have given explicit proofs of two conjectural closed-form expressions for the Poincar\'e polynomials of generalized GHZ and W states, which were proposed in \cite{Mainiero:2019enr}. Second, we have expanded the class of local unitary invariants that can be computed using entanglement $k$-forms. In addition to the dimensions of the cohomology groups, which are invariants that capture some information about the entanglement structure of a state, we have also defined the spectrum of the entanglement Laplacian and a collection of intersection numbers. Both of these new invariants are motivated by analogous geometrical quantities and capture additional data about entanglement in pure states, such as Schmidt coefficients across bipartitions in the case of certain intersection numbers. We have also studied these new invariants numerically in several examples.

There are many interesting directions for future investigation. One is to study entanglement cohomology for other classes of states, such as the ground states of quantum spin systems.\footnote{Some results motivated by this direction will appear in \cite{toappear}.} It would be very interesting if the structure of some unitary invariants constructed from entanglement $k$-forms admitted an interpretation in terms of physical data of such systems, such as the Hamiltonian defining the model. For instance, one could study free chiral fermions on a periodic chain, the SYK model, or the transverse field Ising model. Perhaps the proof technique developed in this work to establish Conjectures \ref{GHZ_conj} and \ref{W_conj} could also be applied to prove exact expressions for the Poincar\'e polynomials of some such models, at least in certain special cases.

Another avenue for research is a more systematic investigation of the invariant theory of the local unitary invariants that we have constructed here. We have not attempted to find functional relations between, for instance, the data in the spectrum of the entanglement Laplacian and the intersection numbers, but it is possible that these LU invariants are dependent in some cases with party number $n \geq 3$. For instance, one might seek such relations using machine learning, e.g. by training a neural network to predict a subset of our LU invariants from the others, and investigate the model's performance. Even if one were to sharply characterize a maximal independent subset of all the LU invariants that have been constructed from entanglement cohomology, we have commented that it seems unlikely that it would be complete. There have been many works studying various LU invariants that can be constructed from quantum states -- see, for instance, the incomplete sampling of works \cite{PhysRevLett.83.243,Makhlin:2000jhd,Sudbery:2001jxc} and references therein -- and a complete classification of these invariants seems quite challenging. It would be interesting if enriching the formalism of entanglement cohomology with some additional structure, and using this to construct and categorize still more LU invariants, might facilitate further progress on this problem.

Finally, it would be quite intriguing to see whether entanglement cohomology could be generalized to characterize either entanglement in more general quantum systems (such as quantum field theories) or to describe other properties of quantum models. We have remarked that one of the fundamental objects in our construction, the coboundary operator (\ref{d_defn}), essentially takes the form of the simplicial coboundary associated with the $(n-1)$ simplex of subsystems, but with coefficients that take values in endomorphisms of certain reduced density matrices, along with an appropriate ``twisting'' by an inclusion map. It is conceivable that a similar structure with coefficients that take values in a different algebra of operators might be used to classify other resources in quantum systems, such as magic \cite{PhysRevA.71.022316,Emerson:2013zse,PhysRevLett.118.090501}. One could also replace the role of the $(n-1)$-simplex in this construction with other objects and investigate applications of the resulting formalism. We leave this important direction to future work.

\section*{Acknowledgements}

We are very grateful to Ning Bao for an authentic shockwave of helpful discussions on the subject of this article. C.\,F. would like to thank Tom Mainiero for many insightful comments and early collaboration on work related to the entanglement Laplacian. C.\,F. would also like to acknowledge Kasra Mossayebi and Gregor Sanfey for fruitful conversations, feedback on this manuscript, and collaboration on related topics. C.\,F. is supported by the National Science Foundation under Cooperative Agreement PHY-2019786 (the NSF AI Institute for Artificial Intelligence and Fundamental Interactions). K.\,F. acknowledges support from Professor Ning Bao at Northeastern University.

\appendix

\section{Review of Simplicial Homology}\label{app:simp}

As we have described in the body of this article, entanglement cohomology is closely related to ordinary simplicial cohomology, although with coefficients that take values in the endomorphism algebras of reduced density matrices. The proofs of the forms of the Poincar\'e polynomials for generalized GHZ and W states in Section \ref{sec:proofs} simply reduce the computation of cohomology groups to arguments about the $(n-1)$-simplex. In particular, the primary fact which we use to prove Lemma \ref{simplex_lemma} is that the $(n-1)$-simplex is contractible. Here we collect a self-contained review of this statement and related definitions in simplicial homology (dual statements apply to cohomology). This appendix contains only standard material that can be found in any textbook on algebraic topology, such as \cite{hatcher2002algebraic}.

It is convenient to introduce some terminology. Recall that a \emph{homotopy} between a pair of continuous functions $f, g : X \to Y$ from a topological space $X$ to another topological space $Y$ is a continuous function $F : X \times [0 , 1] \to Y$ such that $F ( x, 0 ) = f ( x )$ and $F ( x , 1 ) = g ( x )$ for all $x \in X$. A subspace $Y$ of a topological space $X$ is called a \emph{deformation retract} of $X$ if there exists a continuous function $F : X \times [ 0 , 1 ] \to X$ such that, for all $x \in X$ and $y \in Y$, we have $F ( x, 0 ) = x$, $F ( x, 1 ) \in Y$, and $F ( y, 1 ) = y$. The function $F$ is a homotopy from the identity map on $X$ to a function $f(x) = F ( x, 1 )$ which is the identity map on $Y$. Intuitively, such a homotopy continuously deforms the larger space $X$ onto its subspace $Y$. A space $X$ is said to be \emph{contractible} if the identity map on $X$ is homotopic to a constant map, or equivalently, if there exists a deformation retract from $X$ to a point. A contractible space $X$ has trivial homology groups $H_k ( X )$ for all $k > 0$.

To build intuition, it may be useful to consider the \emph{geometric} $n$-simplex, defined by
\begin{align}\label{geometric_n_simplex}
    \Upsilon^n = \left\{ ( t_0 , \ldots, t_n ) \in \mathbb{R}^{n+1} \mid t_i \geq 0 \, , \, \sum_{i=0}^{n} t_i = 1 \right\} \, .
\end{align}
This subset of $\mathbb{R}^{n+1}$ is simply the convex hull of the unit vectors $e_0 = ( 1 , 0 , \ldots, 0 )$, $e_1 = ( 0 , 1 , 0 , \ldots, 0)$, $\ldots$, $e_n = ( 0 , \ldots, 0, 1 )$. It is easy to see why the $n$-simplex (\ref{geometric_n_simplex}) is contractible. Fix a point $p \in \Upsilon^n$ and define the homotopy $F : \Upsilon^n \times [ 0 , 1 ] \to \Upsilon^n$ by
\begin{align}\label{homotopy_example}
    F ( x, s ) = ( 1 - s ) x + s p \, .
\end{align}
We see that $F$ is a continuous function of its arguments, and for any $x \in \Upsilon^n$ and $s \in [0 , 1]$, the quantity $F ( x, s )$ belongs to $\Upsilon^n$ because it is a convex combination of two points in a convex set. Furthermore, one has $F ( x, 0 ) = x$ and $F ( x, 1 ) = p$, so the map $F$ shows that the point $p \in \Upsilon^n$ is a deformation retract of the entire space $\Upsilon^n$ onto a point.\footnote{The same proof applies for any convex subset $X \subset \mathbb{R}^n$, since for any such space the quantity on the right side of (\ref{homotopy_example}) belongs to $X$ by the convexity assumption, so all convex sets are contractible.} This establishes that the $n$-simplex $\Upsilon^n$ is contractible and thus $H_k ( \Upsilon^n )$ is trivial for $k > 0$.

Let us now see how to rephrase this argument somewhat more abstractly in the language of simplicial homology. An \emph{abstract simplicial complex} $K$ is a set of vertices $V$ along with a set of finite subsets of $V$, called \emph{simplices}, such that if $\sigma \in K$ and $\tau \subseteq \sigma$ then $\tau \in K$. For instance, let $V = \{ 1, 2, 3 \}$ and suppose that we form the complete simplicial complex which includes all simplices that can be built from $V$.
\begin{enumerate}[label = (\alph*)]
    \item The $0$-simplices $\{ 1 \}$, $\{ 2 \}$, and $\{ 3 \}$ are points.

    \item The $1$-simplices $\{1 , 2 \}$, $\{1, 3 \}$, and $\{2, 3 \}$ are line segments or edges.

    \item\label{two_simplex} The $2$-simplex $\{ 1, 2, 3 \}$ is a triangle, including its interior, with the given vertices.
\end{enumerate}
Every abstract simplicial complex of this form can be realized as a geometric simplicial complex in some $\mathbb{R}^N$. The subset $\Upsilon^n$ of $\mathbb{R}^{n+1}$ in equation (\ref{geometric_n_simplex}) is an example of a geometric simplicial complex.

Given a simplicial complex $K$ with $n$ vertices, we can build a chain complex whose homology is called the \emph{simplicial homology} of the simplicial complex. Fix an ordering of the vertices of $K$ and write each $k$-simplex as a list of vertices in square brackets, like
\begin{align}
    [ v_0 , \ldots, v_k ] \, , \qquad v_0 < v_1 < \ldots < v_k \, .
\end{align}
The space $C_k ( K )$ of \emph{k-chains} consists of finite linear combinations of such lists of vertices,
\begin{align}
    c = \sum_i a_i [ v_0^{(i)} , \ldots , v_k^{(i)} ] \, ,
\end{align}
where here we take $a_i \in \mathbb{Z}$, and by convention we choose $C_k ( K ) = 0$ if $k < 0$ or $k > n$. One can also allow the coefficients $a_i$ to take values in other fields, like $\mathbb{R}$ or $\mathbb{C}$.

There is an induced ordering on the collection of vertices in any $k$-chain due to the ordering of the vertices themselves, and we choose to define a $k$-chain with opposite ordering to be the negative of its ordered version. For instance,
\begin{align}
    [ v_1 , v_0 ] = - [ v_0 , v_1 ] \, .
\end{align}
The \emph{boundary} operator
\begin{align}
    \partial_k : C_k \to C_{k-1}
\end{align}
is defined on any $k$-chain by
\begin{align}
    \partial_k [ v_0 , \ldots, v_k ] = \sum_{i=0}^{k} ( - 1 )^i [ v_0 , \ldots , \widehat{v}_i , \ldots, v_k ] \, ,
\end{align}
where as usual the hat indicates that this vertex is omitted from the list.

We can then define homological notions in the usual way. Elements of $C_k$ which are annihilated by $\partial_k$ belong to
\begin{align}
    Z_k = \mathrm{ker} ( \partial_k ) 
\end{align}
and are called \emph{cycles}. Elements of $C_k$ which are themselves boundaries belong to
\begin{align}
    B_k = \mathrm{im} ( \partial_{k+1} )
\end{align}
and are thus called \emph{boundaries}. The definition of the boundary operator ensures that $\partial_k \circ \partial_{k+1} = 0$, which we sometimes write as $\partial^2 = 0$, suppressing indices. In particular, $B_k \subset Z_k$, so we can define the quotient as the \emph{k-th homology group}
\begin{align}
    H_k = \frac{Z_k}{B_k} \, .
\end{align}
Let us now see how a homotopy like (\ref{homotopy_example}) is encoded at the level of this chain complex. We write $C_\bullet = \left( \ldots \to C_{k+1} \overset{\partial_{k+1}}{\to} C_k \overset{\partial_{k}}{\to} C_{k-1} \to \ldots \right)$ to represent an abstract chain complex. A \emph{chain map} $f : C_\bullet \to C_\bullet$ is a collection of functions $f_k : C_k \to C_k$ which commutes with the boundary operation in the sense that
\begin{align}
    \partial_k \circ f_k = f_{k-1} \circ \partial_k \, \text{ for all } k \, .
\end{align}
Given two such chain maps $f, g : C_\bullet \to C_\bullet$, a \emph{chain homotopy} $h$ between $f$ and $g$ is a collection of maps $h_k : C_k \to C_{k+1}$ such that
\begin{align}
    f_k - g_k = \partial_{k+1} h_k + h_{k-1} \partial_k
\end{align}
for each $k$. We suppress indices and write this formula as a single identity for the complex,
\begin{align}
    f - g = \partial h + h \partial \, .
\end{align}
The existence of such a chain homotopy can often be used to show that certain homology groups must vanish. At the level of abstract chain complexes -- which need not yet correspond to the simplicial complex of a space --  suppose that there exists a chain homotopy $h$ between $f = \mathrm{id}$ and $g = 0$. This means that
\begin{align}\label{chain_homotopy_identity}
    \mathrm{id} = \partial h + h \partial \, .
\end{align}
We wish to show that the homology of the chain complex is trivial in positive degrees, which means that every cycle is a boundary. Suppose that $z \in Z_k$ so that $\partial z = 0$. Using the identity (\ref{chain_homotopy_identity}), one finds that
\begin{align}\label{trivial_homology_one}
    z = \left( \partial h + h \partial \right) z = \partial h z + h \partial z = \partial ( h z ) \, ,
\end{align}
where we have used $\partial z = 0$ to drop the second term in the identity. But then the result $z = \partial ( h z )$ immediately implies that $z \in B_k$ is a boundary. So we conclude that
\begin{align}
    H_k = \frac{Z_k}{B_k} = 0 \, ,
\end{align}
and the chain complex has trivial homology groups $H_k$ for $k > 0$.

We can now apply this general algebraic result to the specific case of the complete $n$-simplex $\Upsilon^n$.\footnote{Although the complete $n$-simplex which includes all subsets of the vertex set is contractible, more general simplicial complexes with some simplices omitted can have non-trivial homology. For instance, omitting the $2$-simplex \ref{two_simplex} in the example above gives a triangle without its interior, which has $H_1 \cong \mathbb{Z}$.} To demonstrate contractibility, we must explicitly construct a chain homotopy $h$ satisfying (\ref{chain_homotopy_identity}) for the chain complex of the simplex. This algebraic operator is the direct analogue of the geometric contraction to a point $p$ defined in equation (\ref{homotopy_example}).

Let $K$ be the simplicial complex of the $n$-simplex with vertices $\{0, 1, \ldots, n\}$. We fix the vertex $0$ to play the role of the contraction point (the apex of the cone).\footnote{This argument is closely related to the comments, around equation (\ref{solved_constraint}), that every collection of variables $(x_I)$ obeying (\ref{alternating_constraint}) is uniquely determined by its values on subsets containing a fixed single element of $V_n$; that fixed element plays the role of the contraction point.} For any $k$-simplex $\sigma = [v_0, \ldots, v_k]$, define $h_k : C_k \to C_{k+1}$ by prepending $0$ to the list of vertices:
\begin{align}
    h_k ( [ v_0 , \ldots, v_k ] ) = [ 0, v_0 , \ldots, v_k ] \, .
\end{align}
Note that if the vertex $0$ is already present in the simplex (i.e., if $v_0 = 0$), the resulting symbol $[0, 0, v_1, \ldots, v_k]$ contains a repeated vertex and is therefore treated as zero in the chain complex, consistent with the alternating property of chains.

We now verify that this map $h$ is the required chain homotopy between the identity map and the zero map in positive degrees. Let us compute the boundary of the chain $h(\sigma)$ for a $k$-simplex $\sigma = [v_0, \ldots, v_k]$:
\begin{align}
    \partial_{k+1} ( h_k ( \sigma ) ) &= \partial_{k+1} [ 0, v_0 , \ldots, v_k ] \nonumber \\
    &= [ v_0 , \ldots, v_k ] + \sum_{i=0}^{k} ( - 1 )^{i+1} [ 0 , v_0 , \ldots, \widehat{v}_i , \ldots, v_k ] \, .
\end{align}
The first term arises from omitting the first vertex (the new vertex $0$), which comes with a sign factor $(-1)^0 = +1$. The subsequent terms arise from omitting the vertices $v_i$. In the new list $[0, v_0, \ldots]$, the vertex $v_i$ is at position $i+1$, so the sign factor is $(-1)^{i+1}$.

We can rewrite the sum in the second line by factoring out a minus sign and recognizing the action of the map $h$ and the original boundary operator $\partial_k$:
\begin{align}
    \sum_{i=0}^{k} ( - 1 )^{i+1} [ 0 , v_0 , \ldots, \widehat{v}_i , \ldots, v_k ] &= - h_{k-1} \left( \sum_{i=0}^{k} ( - 1 )^{i} [ v_0 , \ldots, \widehat{v}_i , \ldots, v_k ] \right) \nonumber \\
    &= - h_{k-1} \left( \partial_k [ v_0 , \ldots, v_k ] \right) \nonumber \\
    &= - ( h \partial ) ( \sigma ) \, .
\end{align}
Combining these results, we find that for any $k$-simplex $\sigma$,
\begin{align}
    \partial ( h \sigma ) = \sigma - h ( \partial \sigma ) \, .
\end{align}
Rearranging terms, we obtain the identity
\begin{align}
    \partial h + h \partial = \mathrm{id} \, .
\end{align}
This holds for all chains in degrees $k > 0$. By the logic around equation (\ref{chain_homotopy_identity}), this implies that for every cycle $z \in Z_k$ with $k > 0$, we have $z = \partial ( h z )$, and therefore $z$ is a boundary.

Consequently, $H_k ( \Upsilon^n ) = 0$ for all $k > 0$. This confirms that the $n$-simplex is algebraically contractible, justifying the arguments used in the proof of Lemma \ref{simplex_lemma}.

\bibliographystyle{JHEP}
\bibliography{main}

\end{document}